\documentclass[
reprint,
superscriptaddress,
frontmatterverbose, 
showpacs,
preprintnumbers,
nofootinbib,
amsmath,
amssymb,
aps,
floatfix,
twocolumn
]{revtex4-2}

\usepackage{graphicx}
\usepackage{dcolumn}
\usepackage{bm}
\usepackage{hyperref}
\usepackage{xcolor}
\usepackage{subcaption}
\usepackage{booktabs,array}

\begin{document}


\title{
Boson--Fermion Atomic Dark Matter  
}

\author{Natsumi~Nagata}
 \email{natsumi@hep-th.phys.s.u-tokyo.ac.jp}
  \affiliation{Department of Physics, University of Tokyo, Bunkyo-ku, Tokyo
 113--0033, Japan
 }
\author{Tsutomu T. Yanagida}
 \affiliation{Kavli Institute for the Physics and Mathematics of the Universe (WPI), UTIAS, The University of Tokyo, Chiba 277-8583, Japan
}

\date{\today}

\begin{abstract}

  We propose a simple scenario of boson--fermion atomic dark matter, in which the dark matter abundance originates from a primordial particle--antiparticle asymmetry. The dark sector consists of a complex scalar and a Dirac fermion carrying opposite charges under an unbroken $\mathrm{U}(1)_D$ gauge symmetry. Their asymmetries are generated non-thermally through the CP-violating decays of heavy Majorana fermions produced by inflaton decay, in close analogy with non-thermal leptogenesis in the visible sector. The symmetric components efficiently annihilate into massless dark photons, while the surviving scalar and fermion populations cannot annihilate with each other and instead recombine into neutral boson--fermion dark atoms. Remarkably, the same interactions responsible for generating the primordial asymmetry induce atom--antiatom oscillations after recombination, so that the present-day dark matter can be approximately symmetric despite its asymmetric origin. Residual ionized components or sufficiently extended dark atoms can also give rise to appreciable dark-matter self-interactions, potentially leaving observable imprints on the formation and structure of dark matter halos and offering complementary cosmological and astrophysical probes of the dark sector. The massless dark photon survives as dark radiation, yielding a characteristic contribution $\Delta N_{\rm eff}\simeq0.05$--$0.07$ for the thermal history considered here. This prediction lies within the reach of future precision CMB observations, providing a direct cosmological test of the scenario.

\end{abstract}

\maketitle


\section{Introduction}
\label{sec:introduction}

The matter--antimatter asymmetry is essential for the existence of the visible Universe. Without an excess of matter over antimatter in the early Universe, almost all particles and antiparticles would have annihilated away, leaving behind a Universe drastically different from the one we observe today. This profound role of the matter--antimatter asymmetry naturally raises the possibility that an analogous asymmetry may also be present in the dark sector. In particular, the observed dark matter abundance may originate, at least in part, from an excess of dark matter (DM) particles over their antiparticles. This idea, broadly known as asymmetric dark matter~\cite{Hut:1979xw, Nussinov:1985xr, Gelmini:1986zz, Barr:1990ca, Barr:1991qn, Kaplan:1991ah, Dodelson:1991iv, Kuzmin:1996he, Hooper:2004dc, Kitano:2004sv, Kitano:2005ge, Farrar:2005zd, Kaplan:2009ag, Ibe:2011hq, Nagata:2016knk}, provides an intriguing framework in which the cosmic abundances of visible and dark matter may share a common underlying origin.

Motivated by these considerations, in this work we propose a simple realization of asymmetric dark matter with rich and testable phenomenology. We assume that the inflaton decays not only into right-handed neutrinos but also into a new set of singlet Majorana fermions. The non-thermally produced right-handed neutrinos subsequently undergo out-of-equilibrium decays, generating the baryon asymmetry of the visible sector through non-thermal leptogenesis~\cite{Fukugita:1986hr,Lazarides:1990huy,Kumekawa:1994gx,Asaka:1999yd,Asaka:1999jb}. In close analogy, the new Majorana fermions decay out of equilibrium into a singlet complex scalar and a singlet Dirac fermion, generating a dark-sector asymmetry with equal numbers of the two species. A distinctive feature of this setup is that the asymmetry is shared between a boson and a fermion. Since these are distinct particles with different spins, they cannot annihilate with each other and thereby erase the asymmetry. This mechanism was originally proposed in Ref.~\cite{Nagata:2026pbj}. Both the scenario of Ref.~\cite{Nagata:2026pbj} and the present one feature a dark $\mathrm{U}(1)_D$ gauge symmetry; the crucial difference is that Ref.~\cite{Nagata:2026pbj} considered a spontaneously broken $\mathrm{U}(1)_D$, whereas here we assume that it remains exact and unbroken. The corresponding gauge boson is therefore massless, giving rise to long-range interactions in the dark sector.

The symmetric components of the scalar and fermion efficiently annihilate into the massless $\mathrm{U}(1)_D$ gauge bosons, which we refer to as dark photons, leaving behind the asymmetric populations generated in the decays of the heavy Majorana fermions. The surviving scalar and fermion then experience an attractive dark Coulomb force and form electrically neutral \textit{boson--fermion dark atoms}, which constitute the dark matter in our scenario.\footnote{For other realizations of atomic dark matter considered in the literature, see, e.g., Refs.~\cite{Kaplan:2009de, Khlopov:2010ik, Kaplan:2011yj,Cyr-Racine:2012tfp}.} As we will show, the presence of an exact dark gauge symmetry and the resulting atomic structure lead to a rich variety of cosmological and astrophysical phenomena, making this scenario qualitatively different from the broken-$\mathrm{U}(1)_D$ realization considered in Ref.~\cite{Nagata:2026pbj}.


\begin{figure*}
    \centering
    \includegraphics[width=0.9\textwidth]{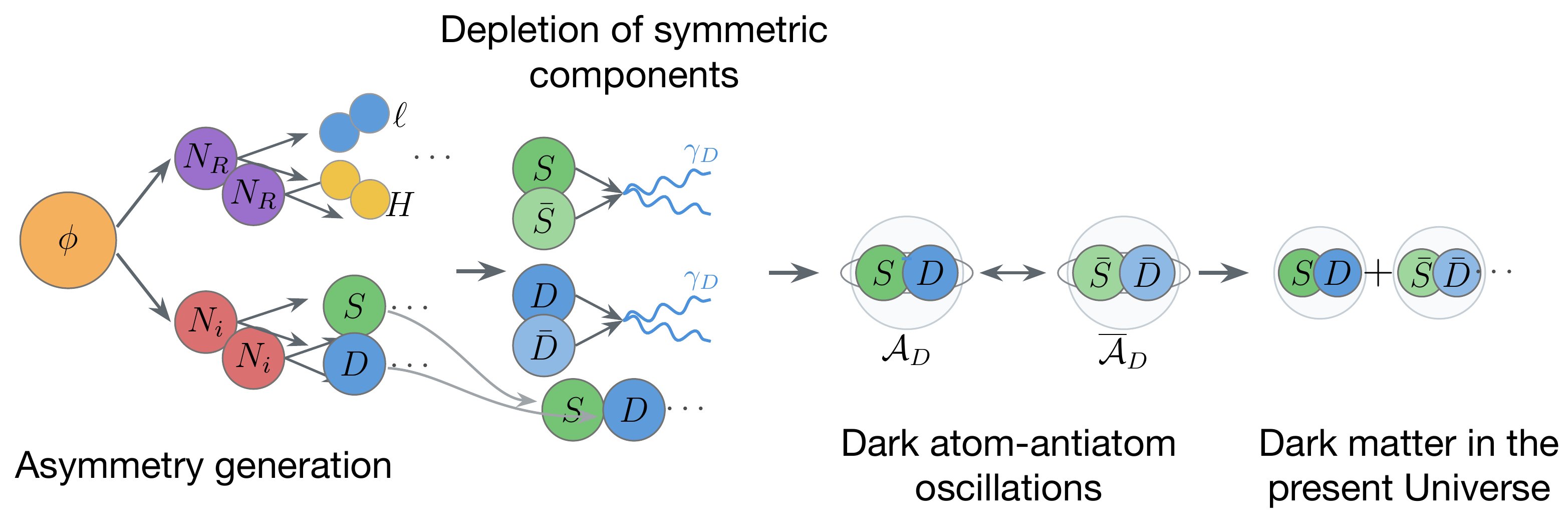}
    \caption{
    Schematic illustration of the boson--fermion atomic dark matter scenario.
    }
    \label{fig:schematic}
\end{figure*}

After dark-atom formation, atom--antiatom oscillations convert the initially asymmetric population into a symmetric mixture of atoms and antiatoms~\cite{Buckley:2011ye, Cirelli:2011ac, Tulin:2012re, Okada:2012rm}.\footnote{For an example of particle--antiparticle oscillations in composite asymmetric dark matter, see Ref.~\cite{Ibe:2019yra}.} Thus, remarkably, the dark matter in the present Universe is symmetric despite having originated from an asymmetric production mechanism. Figure~\ref{fig:schematic} schematically illustrates the boson--fermion atomic dark matter scenario proposed in this work. Meanwhile, the relic massless dark photons remain as dark radiation and contribute \(\Delta N_{\rm eff}\simeq 0.05\text{--}0.07\). This characteristic prediction lies within the reach of future precision measurements of the cosmic microwave background~\cite{Baumann:2017gkg}, providing a direct cosmological test of the scenario. Residual ionized components or sufficiently extended dark atoms can also give rise to appreciable dark-matter self-interactions, potentially leaving observable imprints on the formation and structure of dark matter halos and offering complementary cosmological and astrophysical probes of the dark sector.

The remainder of this paper is organized as follows. In Sec.~\ref{sec:model}, we introduce the model and specify the particle content, symmetries, and interactions of the dark sector. In Sec.~\ref{sec:asymmetry}, we discuss the non-thermal generation of the dark-sector asymmetry through the decays of the heavy Majorana fermions, as well as the washout processes that can affect the resulting asymmetry. In Sec.~\ref{sec:depletion}, we study the annihilation of the symmetric components of the dark scalar and fermion into dark photons. In Sec.~\ref{sec:bfda}, we introduce the boson--fermion dark atom and discuss its basic properties and relic abundance. In Sec.~\ref{sec:dr}, we examine the thermal decoupling of the visible and dark sectors and evaluate the resulting dark-photon contribution to \(\Delta N_{\rm eff}\). In Sec.~\ref{sec:recombination}, we discuss dark recombination, estimate the residual ionized fraction, and comment on its possible phenomenological consequences. In Sec.~\ref{sec:oscillation}, we study dark atom--antiatom oscillations and their impact on the late-time dark matter population. Finally, Sec.~\ref{sec:discussion} is devoted to discussion and conclusions.

\section{Model}
\label{sec:model}

We consider a dark sector with an unbroken Abelian gauge symmetry \(\mathrm{U}(1)_D\) and a corresponding massless dark photon, denoted by \(\gamma_D\). Throughout this paper, all fermionic fields are written as left-handed two-component Weyl spinors. The dark sector contains a complex scalar \(S\), a vector-like pair of Weyl fermions \(D\) and \(D^c\), and heavy gauge-singlet Majorana Weyl fermions \(N_i\), where \(i\) labels generations. The pair \(D\) and \(D^c\) carries opposite \(\mathrm{U}(1)_D\) charges and forms a vector-like Dirac fermion. All of these new fields are singlets under the Standard Model (SM) gauge group \(G_{\rm SM}\).

The fields $N_i$ are assumed to be distinct from the right-handed neutrinos $N_R^c$ associated with the neutrino sector. This separation can be realized by means of the \(B-L\) symmetry. We assume that $\mathrm{U}(1)_{B-L}$ is spontaneously broken at a high scale, leaving a remnant discrete symmetry~\cite{Krauss:1988zc, Ibanez:1991hv, Ibanez:1991pr, Martin:1992mq}. A convenient realization is the usual matter parity, \(P_{B-L}=(-1)^{3(B-L)}\). The dark-sector fields $N_i$, $D$, $D^c$, and $S$ are taken to have vanishing $B-L$ charge and are therefore even under $P_{B-L}=(-1)^{3(B-L)}$, whereas the SM matter fields, including $N_R^c$, are odd. The Higgs doublet $H$ is even. This assignment prevents the heavy singlets $N_i$ from mixing or directly interacting with the SM neutrino sector through renormalizable operators, \textit{e.g.}, \(N_i N_R^c\) and \(N_i L H\). In addition, the couplings between the right-handed neutrinos and the dark-sector matter fields, \(S D N_R^c\) and \(S^\dagger D^c N_R^c\), are forbidden.

The relevant quantum numbers are summarized in Table~\ref{tab:charges}.

\begin{table}[ht]
\centering
\renewcommand{\arraystretch}{1.2}
\begin{tabular}{c c c c c}
\toprule
Field & Spin & $G_{\rm SM}$ & $\mathrm{U}(1)_D$ & $P_{B-L}$ \\
\midrule
$N_i$   & $1/2$ & $(\bm{1},\bm{1},0)$ & $0$  & $+$ \\
$D$     & $1/2$ & $(\bm{1},\bm{1},0)$ & $+1$ & $+$ \\
$D^c$   & $1/2$ & $(\bm{1},\bm{1},0)$ & $-1$ & $+$ \\
$S$     & $0$   & $(\bm{1},\bm{1},0)$ & $-1$ & $+$ \\
\midrule
$N_R^c$   & $1/2$ & $(\bm{1},\bm{1},0)$ & $0$  & $-$ \\
$L$     & $1/2$ & $(\bm{1},\bm{2},-1/2)$ & $0$ & $-$ \\
$H$     & $0$   & $(\bm{1},\bm{2},+1/2)$ & $0$ & $+$ \\
\bottomrule
\end{tabular}
\caption{Field content and charge assignments.}
\label{tab:charges}
\end{table}

The Lagrangian terms relevant for our discussion are
\begin{align}
\mathcal{L}
=&-\biggl[
\frac12 M_{ij}N_iN_j
+m_D D D^c \nonumber \\
&+g_i N_i D S
+\widetilde g_i N_i D^c S^\dagger
+\text{h.c.}
\biggr] \nonumber \\
&- m_S^2 |S|^2
-\lambda_{HS} |H|^2 |S|^2
-\frac{\epsilon}{2} B_{\mu\nu} F_D^{\mu\nu} \,,
\label{eq:massYuk}
\end{align}
where \(F_{D \mu \nu}\) and \(B_{\mu\nu}\) denote the field-strength tensors of the $\mathrm{U}(1)_D$ gauge field and the hypercharge gauge field, respectively, and $M_{ij}=M_{ji}$ is the Majorana mass matrix of the heavy singlets \(N_i\). We have omitted renormalizable terms that are not directly relevant for the discussion below, such as the dark-scalar quartic interaction \(|S|^4\).

We assume that the scalar potential is such that $S$ does not acquire a vacuum expectation value (VEV), so that $\mathrm{U}(1)_D$ remains unbroken and the dark photon is exactly massless. 

The last two terms in Eq.~\eqref{eq:massYuk} provide possible portals between the visible and dark sectors.  The Higgs portal coupling \(\lambda_{HS}\) can bring the scalar \(S\) into thermal equilibrium with the SM plasma and thus plays an important role in determining the thermal history of the dark sector.  As we discuss below, the resulting thermal contact and subsequent decoupling determine the dark-sector temperature and are therefore subject to cosmological constraints.  In contrast, we assume the gauge kinetic mixing\footnote{Mixing between neutral vector fields was considered early on in the context of vector-meson dominance~\cite{Sakurai:1960ju} and was formulated in a gauge-invariant Lagrangian framework in Ref.~\cite{Kroll:1967it}.  Kinetic mixing between two Abelian gauge fields and its radiative generation in theories containing particles charged under both gauge groups were discussed in Ref.~\cite{Holdom:1985ag}; see also Ref.~\cite{Fabbrichesi:2020wbt} for further discussions and reviews.  For an analogous treatment of scalar kinetic mixing, see Ref.~\cite{Fukuda:1974kn}.} parameter \(\epsilon\) to be negligibly small in the following discussion.  This can naturally be assured if hypercharge \(\mathrm{U}(1)_Y\) is embedded into a simple non-Abelian grand unified gauge group and no states carrying both hypercharge and \(\mathrm{U}(1)_D\) charge are present below the unification scale.  In this case, the renormalizable kinetic-mixing operator is forbidden above the GUT-breaking scale, and its radiative generation at lower energies can be strongly suppressed~\cite{Gherghetta:2019coi}.\footnote{For the generation of kinetic mixing in grand-unified constructions, see, e.g., Refs.~\cite{Kamada:2019jch, Gherghetta:2019coi}.}

\section{Asymmetry in the dark sector}
\label{sec:asymmetry}

We now discuss how a particle--antiparticle asymmetry is generated in the dark sector and how it subsequently evolves. The mechanism closely parallels leptogenesis~\cite{Fukugita:1986hr}. This mechanism for asymmetric dark matter was originally proposed in Ref.~\cite{Nagata:2026pbj}. Related scenarios in which heavy Majorana fermion decays generate a dark-sector asymmetry have been studied in Refs.~\cite{An:2009vq,Falkowski:2011xh,Josse-Michaux:2011sjn,Narendra:2018vfw}.

We assume that both the heavy Majorana fermions $N_i$ and the right-handed neutrinos in the SM sector are produced non-thermally through inflaton decays. The heavy Majorana fermions \(N_i\) subsequently decay into dark-sector matter fields through the Yukawa couplings $g_i$ and $\widetilde g_i$ in Eq.~\eqref{eq:massYuk}. Because $N_i$ are Majorana fermions, they can decay into both particle and antiparticle final states. Interference between tree-level and loop amplitudes can generate a CP asymmetry between these decay channels, producing particle--antiparticle asymmetries in the dark sector. This mechanism is a straightforward extension of standard non-thermal leptogenesis~\cite{Fukugita:1986hr}, in which inflaton decays produce heavy Majorana fermions whose subsequent CP-violating decays generate a cosmological asymmetry~\cite{Lazarides:1990huy,Kumekawa:1994gx,Asaka:1999yd,Asaka:1999jb}.

To track the particle--antiparticle asymmetry that determines the late-time dark matter abundance, we introduce a quantum number $X$. We assign \(X(D)=X(S)=+1\), \(X(D^c)=X(S^\dagger)=-1\), and \(X(N_i)=0\). The Dirac mass term $m_D D D^c$, dark gauge interactions, Higgs portal interaction, and kinetic mixing all conserve $X$. By contrast, the interactions $g_iN_iDS$ and $\widetilde g_iN_iD^cS^\dagger$ violate $X$ by $+2$ and $-2$, respectively. Thus, $X$ is not an exact symmetry while the $N_i$-mediated interactions are active, but becomes an approximate symmetry after the heavy Majorana fermions decouple.

We define the asymmetry yields as
\begin{align}
Y_{\Delta D}
&\equiv \frac{n_{\Psi_D}-n_{\overline{\Psi}_D}}{s}\,,
\qquad
Y_{\Delta S}
\equiv \frac{n_S-n_{S^\dagger}}{s}\,,
\label{eq:ydels}
\end{align}
where $\Psi_D$ is the Dirac fermion formed by $D$ and $(D^c)^\dagger$, and $s$ is the entropy density. Assuming that the Universe has no net dark gauge charge after inflation, the decays of $N_i$ generate equal asymmetries in the fermion and scalar sectors, \(Y_{\Delta D}=Y_{\Delta S}\equiv Y_\Delta\). The total $X$ asymmetry is therefore $Y_X=2Y_\Delta$. After dark recombination~\cite{Kaplan:2009de, Kaplan:2011yj}, each dark atom contains one Dirac fermion and one scalar. The dark-atom yield is thus $Y_{\mathcal A_D}=Y_\Delta$, provided that the symmetric components have been efficiently depleted and residual ionization and subsequent reannihilation can be neglected. 

Without loss of generality, we work in a basis in which the Majorana mass matrix is real and diagonal, \(M_{ij}=M_i\delta_{ij}\), with \(M_i>0\). We define the CP asymmetry per $N_i$ decay as the net excess of Dirac particles, or equivalently of $S$ particles, produced per decay:
\begin{align}
\epsilon_i
\equiv
\frac{1}{\Gamma_i}
\biggl[
&
\Gamma(N_i\to DS)
+\Gamma(N_i\to D^{c\dagger}S)
\nonumber\\
&-\Gamma(N_i\to D^\dagger S^\dagger)
-\Gamma(N_i\to D^cS^\dagger)
\biggr]\,,
\label{eq:epsilon-def}
\end{align}
where \(\Gamma_i\) is the total tree-level decay width of \(N_i\),
\begin{align}
\Gamma_i
\equiv
\frac{M_i}{16\pi}
\left(
|g_i|^2+|\widetilde g_i|^2
\right).
\label{eq:Ni-width}
\end{align}
The asymmetry vanishes at tree level and arises from the interference of the tree-level decay amplitudes with the one-loop vertex and self-energy corrections. A nonzero asymmetry therefore requires at least two heavy Majorana fermions \(N_i\) and irreducible complex phases in their Yukawa couplings. Indeed, with only a single heavy Majorana fermion, the relevant phases can be removed by field redefinitions, so that no physical CP-violating invariant contributing to the decay asymmetry remains. With two or more \(N_i\), by contrast, rephasing-invariant CP-odd combinations can be constructed. These include, for example, \(\operatorname{Im}\!\left[(g_i^*g_j)^2\right]\) and \(\operatorname{Im}\!\left[(\widetilde g_i^*\widetilde g_j)^2\right]\) for \(i\neq j\), whose nonzero values signal CP violation and generate the dark-sector asymmetry through the interference terms.

We denote the inflaton by $\phi$, its mass by $m_\phi$, and the reheating temperature by $T_R$. Rather than specifying the inflation sector in detail, we parameterize heavy Majorana fermion production in terms of the inflaton branching fractions, \(B_i^\phi \equiv {\rm Br}(\phi\to N_iN_i)\). Pair production is kinematically allowed for \(m_\phi>2M_i\). As noted above, we also assume that the inflaton decays into right-handed neutrinos in the SM sector, reheating the visible sector and generating the baryon asymmetry of the Universe through non-thermal leptogenesis.

The inflaton decays when the Hubble expansion rate \(H\) becomes comparable to the inflaton decay width \(\Gamma_\phi\): $H\simeq\Gamma_\phi$. We define the reheating temperature through
\begin{align}
\Gamma_\phi
=
H(T_R),
\qquad
H(T)
=
1.66\sqrt{g_*}\frac{T^2}{M_{\rm Pl}},
\label{eq:TR-definition}
\end{align}
where $M_{\rm Pl}=1.22\times10^{19}\ {\rm GeV}$ and \(g_*\) is the effective number of relativistic degrees of freedom. In the instantaneous-reheating approximation, the inflaton abundance at reheating is
\begin{align}
\frac{n_\phi}{s}
\simeq
\frac{3}{4}\frac{T_R}{m_\phi}\,,
\end{align}
where \(n_\phi\) is the inflaton number density.

For simplicity, we focus on the regime \(\Gamma_i \gg \Gamma_\phi\), in which the $N_i$ decay promptly after their production. Since each $\phi\to N_iN_i$ decay produces two $N_i$, their non-thermal yield is
\begin{align}
Y_{N_i}^{\rm NT}
\simeq
\frac{3}{2}
B_i^\phi
\frac{T_R}{m_\phi}\,,
\label{eq:YN-nonthermal}
\end{align}
where \(B_i^\phi ={\rm Br}(\phi\to N_iN_i)\) is the inflaton branching fraction into a pair of \(N_i\). If the produced $N_i$ decay without significant washout or subsequent entropy dilution, the resulting dark-sector asymmetry is
\begin{align}
Y_\Delta^\infty
\simeq
\sum_i
\epsilon_i
Y_{N_i}^{\rm NT}
=
\frac{3}{2}
\frac{T_R}{m_\phi}
\sum_i
B_i^\phi\epsilon_i\,.
\label{eq:Ydelta-nonthermal}
\end{align}

We further assume \(T_R \ll M_i\), so that the inverse decays \(DS \to N_i\) and \(D^c S^\dagger \to N_i\) are exponentially suppressed and the $N_i$ decays remain effectively out of equilibrium.\footnote{The maximum temperature \(T_{\mathrm{max}}\) reached during perturbative reheating can be significantly higher than the conventional reheating temperature~\cite{Chung:1998rq,Giudice:2000ex}. If \(T_{\mathrm{max}}\) is relevant to dark-sector thermalization, the condition above should be imposed using \(T_{\mathrm{max}}\) instead of \(T_R\). The same replacement applies when evaluating the washout condition in Eq.~\eqref{eq:DeltaX-condition}.}

Below the heavy Majorana mass scale, integrating out the $N_i$ generates the dimension-five operators
\begin{align}
{\cal L}_{\rm eff}
\supset
&
\frac{1}{2}C_g(DS)(DS)
+
\frac{1}{2}C_{\widetilde g}
(D^cS^\dagger)(D^cS^\dagger)
\nonumber\\
&+
C_0(DS)(D^cS^\dagger)
+\text{h.c.},
\label{eq:effective-washout}
\end{align}
where
\begin{align}
C_g
&=
\sum_i\frac{g_i^2}{M_i},
&
C_{\widetilde g}
&=
\sum_i\frac{\widetilde g_i^2}{M_i},
&
C_0
&=
\sum_i\frac{g_i\widetilde g_i}{M_i}.
\label{eq:C-coefficients}
\end{align}
The first two operators violate $X$ by four units and induce washout processes such as \(DS \leftrightarrow D^\dagger S^\dagger\) and \(DD \leftrightarrow S^\dagger S^\dagger\), along with the corresponding processes involving $D^c$. The mixed operator proportional to $C_0$, on the other hand, conserves the total $X$ charge. It therefore does not erase the net asymmetry.

The rate of the $X$-violating processes induced by these operators scales as
\begin{align}
\Gamma_{\Delta X}
\simeq
c_{\Delta X}
T^3
\left(
|C_g|^2+|C_{\widetilde g}|^2
\right),
\label{eq:DeltaX-rate}
\end{align}
where $c_{\Delta X}$ is a numerical coefficient of order $10^{-2}$--$10^{-1}$. Since $\Gamma_{\Delta X}/H\propto T$ during radiation domination, the washout constraint is most stringent at the highest temperature reached after the asymmetry is generated. Requiring these $X$-violating processes to remain out of equilibrium therefore yields
\begin{align}
c_{\Delta X}T_R^3
\left(
|C_g|^2+|C_{\widetilde g}|^2
\right)
<
1.66\sqrt{g_*}\frac{T_R^2}{M_{\rm Pl}}.
\label{eq:DeltaX-condition}
\end{align}
Equivalently,
\begin{align}
&\sqrt{|C_g|^2+|C_{\widetilde g}|^2}
\lesssim
1.2\times10^{-12}\ {\rm GeV}^{-1}
\nonumber\\
&\times
\biggl(\frac{c_{\Delta X}}{10^{-2}}\biggr)^{-1/2}
\left(\frac{g_*}{106.75}\right)^{1/4}
\left(\frac{T_R}{10^8\ {\rm GeV}}\right)^{-1/2}.
\label{eq:DeltaX-bound}
\end{align}
This condition is readily satisfied across a broad region of parameter space. For example, couplings of order \(g_i,\widetilde g_i=\mathcal{O}(0.1)\) are compatible with the washout bound for heavy-fermion masses of order \(10^{10}\)--\(10^{11}\ {\rm GeV}\), depending on the number of contributing states and possible cancellations among the terms in Eq.~\eqref{eq:C-coefficients}.

\section{Depletion of symmetric components}
\label{sec:depletion}

We next discuss the depletion of the symmetric components of $D$ and $S$. We assume that the Higgs portal interaction brings the dark and visible sectors into thermal equilibrium at temperatures \(T\gtrsim m_D,m_S\), and that dark gauge interactions maintain internal thermal equilibrium within the dark sector. For the analytic estimates below, we further assume that the two sectors share a common temperature during chemical freeze-out.

As the Universe cools, $D$ and $S$ become non-relativistic and their symmetric components annihilate into dark photons. These annihilations preserve the particle--antiparticle asymmetries generated earlier. Thermal contact between the visible and dark sectors eventually becomes inefficient, after which their comoving entropies are separately conserved. We discuss the subsequent temperature evolution and the dark-photon contribution to $\Delta N_{\rm eff}$ in Sec.~\ref{sec:dr}.

We estimate the depletion using the annihilation channels \(D\overline D \to \gamma_D\gamma_D\) and \(SS^\dagger \to \gamma_D\gamma_D\), where $D$ denotes the Dirac fermion $\Psi_D$ in this section. In the non-relativistic limit, their tree-level cross sections are\footnote{The factor of two in the scalar annihilation cross section follows from scalar QED, including the four-point scalar--photon interaction. The long-range dark force enhances annihilation through the Sommerfeld effect~\cite{Sommerfeld:1931qaf,Hisano:2003ec,Hisano:2004ds, Baldes:2017gzw}. In the Coulomb limit, the enhancement factor is
\begin{align}
{\cal S}(v_{\rm rel})
=
\frac{2\pi\alpha_D/v_{\rm rel}}
{1-\exp(-2\pi\alpha_D/v_{\rm rel})}\,.
\label{eq:Sommerfeld}
\end{align}
Radiative bound-state formation followed by annihilation can further deplete particle--antiparticle pairs at low velocities~\cite{vonHarling:2014kha, Petraki:2015hla}. Neglecting these effects therefore gives a conservative estimate of the dark gauge coupling required to efficiently remove the symmetric component.}
\begin{align}
\left(\sigma v_{\rm rel}\right)_{D\overline D}
=
\frac{\pi\alpha_D^2}{m_D^2}\,,
\qquad
\left(\sigma v_{\rm rel}\right)_{SS^\dagger}
=
\frac{2\pi\alpha_D^2}{m_S^2}\,,
\label{eq:S-annihilation}
\end{align}
where $\alpha_D=g_D^2/(4\pi)$. We retain only these leading $s$-wave contributions in the following estimate.

For each species \(a=D,S\), we define
\begin{align}
Y_a\equiv\frac{n_a}{s}\,,
\quad
Y_{\bar a}\equiv\frac{n_{\bar a}}{s}\,,
\quad
x_a\equiv\frac{m_a}{T}\,,
\end{align}
with \(\bar S=S^\dagger\). We take the common asymmetry to be positive, \(Y_a-Y_{\bar a}=Y_\Delta>0\), and assume that washout and late entropy production are negligible. During radiation domination, the Boltzmann equation for the antiparticle yield is
\begin{align}
\frac{dY_{\bar a}}{dx_a}
=
-\frac{s\langle\sigma v\rangle_a}{Hx_a}
\left[
Y_{\bar a}(Y_{\bar a}+Y_\Delta)
-Y_{a,{\rm eq}}Y_{\bar a,{\rm eq}}
\right]\,,
\label{eq:depletion-boltzmann}
\end{align}
where \(\langle\sigma v\rangle_a\) denotes the thermally averaged annihilation cross section for the corresponding particle--antiparticle pair. The equilibrium yields include the chemical potentials associated with the conserved asymmetry; their product is independent of these chemical potentials.

We approximate the effective numbers of relativistic degrees of freedom as constant over the interval that controls depletion. Using \(s = 2\pi^2g_{*s}T^3/(45)\), we obtain
\begin{align}
\frac{s}{Hx_a}
=
\frac{\lambda_a}{x_a^2}\,,
\quad
\lambda_a
\equiv
0.264\frac{g_{*s}}{\sqrt{g_*}}M_{\rm Pl}m_a\,.
\label{eq:depletion-lambda}
\end{align}
Here, \(g_{*s}\) is defined by the entropy density used to normalize the yields, and \(g_*\) includes all relativistic species contributing to the expansion rate.

After inverse annihilations become negligible, Eq.~\eqref{eq:depletion-boltzmann} can be integrated directly. Introducing the antiparticle-to-particle ratio
\begin{align}
r_a(x_a)\equiv\frac{Y_{\bar a}}{Y_a}
=\frac{Y_{\bar a}}{Y_{\bar a}+Y_\Delta}\,,
\end{align}
we find
\begin{align}
\frac{d\ln r_a}{dx_a}
\simeq
-\frac{\lambda_aY_\Delta
\langle\sigma v\rangle_a}{x_a^2}\,.
\end{align}
Thus, for a constant $s$-wave cross section,
\begin{align}
r_{a,\infty}
&\simeq
r_{a,f}\exp(-A_a)\,,
&
A_a
&\equiv
\frac{\lambda_aY_\Delta
\langle\sigma v\rangle_a}{x_{a,f}}\,,
\label{eq:depletion-ratio}
\end{align}
where \(x_{a,f}=m_a/T_{a,f}\) is a matching point near freeze-out and \(r_{a,f}=r_a(x_{a,f})\). The final yields are consequently
\begin{align}
Y_{\bar a}^{\infty}
&=
\frac{Y_\Delta r_{a,\infty}}
{1-r_{a,\infty}}\,,
&
Y_a^\infty
&=
\frac{Y_\Delta}
{1-r_{a,\infty}}\,.
\label{eq:depletion-final-yields}
\end{align}
These expressions show explicitly that the antiparticle abundance is exponentially suppressed, while the surviving particle yield approaches \(Y_\Delta\).

Within this approximation, setting \(r_{a,f}=1\) gives a conservative estimate of the residual antiparticle fraction, since \(r_{a,f}\leq1\). A sufficient condition for \(r_{a,\infty}<r_{\rm max}\) is therefore
\begin{align}
A_a\gtrsim\ln\frac{1}{r_{\rm max}}\,.
\label{eq:depletion-exponent}
\end{align}
For \(r_{\rm max}=10^{-2}\), this requires \(A_a\gtrsim4.6\). Taking \(x_{a,f}=25\) as a representative matching value gives\footnote{For a more precise calculation, see Ref.~\cite{Graesser:2011wi}.}
\begin{align}
\langle\sigma v\rangle_a
&\gtrsim
1.9\times10^{-25}\ {\rm cm^3\,s^{-1}}
\nonumber\\
&\quad\times
\left(\frac{x_{a,f}}{25}\right)
\left(\frac{g_{*s}/\sqrt{g_*}}{10}\right)^{-1}
\nonumber\\
&\quad\times
\left(\frac{m_a}{300\ {\rm GeV}}\right)^{-1}
\left(\frac{Y_\Delta}{7.3\times10^{-13}}\right)^{-1}.
\label{eq:depletion-cross-section}
\end{align}
The reference asymmetry corresponds to a dark-atom mass of \(600\ {\rm GeV}\) and the observed dark matter abundance, as shown in Eq.~\eqref{eq:Ydelta-observed}. Unlike a universal thermal-relic benchmark, Eq.~\eqref{eq:depletion-cross-section} displays the dependence on the constituent mass and the primordial asymmetry explicitly. In particular, a smaller asymmetry requires a larger annihilation cross section to achieve the same residual antiparticle fraction.

Substituting Eq.~\eqref{eq:S-annihilation}, we obtain
\begin{align}
\alpha_D
&\gtrsim
2.2\times10^{-2}
\left(\frac{x_{D,f}}{25}\right)^{1/2}
\nonumber\\
&\quad\times
\left(
\frac{(g_{*s}/\sqrt{g_*})_{D,f}}{10}
\right)^{-1/2}
\left(\frac{m_D}{300\ {\rm GeV}}\right)^{1/2}
\nonumber\\
&\quad\times
\left(\frac{Y_\Delta}{7.3\times10^{-13}}\right)^{-1/2},
\label{eq:alphaD-D-bound}
\\
\alpha_D
&\gtrsim
1.5\times10^{-2}
\left(\frac{x_{S,f}}{25}\right)^{1/2}
\nonumber\\
&\quad\times
\left(
\frac{(g_{*s}/\sqrt{g_*})_{S,f}}{10}
\right)^{-1/2}
\left(\frac{m_S}{300\ {\rm GeV}}\right)^{1/2}
\nonumber\\
&\quad\times
\left(\frac{Y_\Delta}{7.3\times10^{-13}}\right)^{-1/2},
\label{eq:alphaD-S-bound}
\end{align}
for the fermion and scalar, respectively. The effective degrees of freedom are evaluated at the corresponding matching temperatures. Both conditions must be satisfied to suppress the symmetric components of both species. For \(m_D=m_S=300\ {\rm GeV}\) and the reference asymmetry, the fermion condition is stronger and requires a dark gauge coupling of a few times \(10^{-2}\). Sommerfeld enhancement and bound-state formation followed by annihilation can increase depletion and relax the coupling requirements. As we will see in Sec.~\ref{sec:recombination}, this lower bound is weaker than the lower bound on \(\alpha_D\) required for \(S\) and \(D\) to form dark atoms.

\section{Boson-fermion dark atoms}
\label{sec:bfda}

An important feature of our setup is that \(S\) and \(D\) carry opposite \(\mathrm{U}(1)_D\) charges and therefore experience an attractive Coulomb force mediated by the massless dark photon.  Despite their opposite gauge charges, however, \(S\) and \(D\) do not annihilate with each other.  This is a crucial difference from an ordinary particle--antiparticle pair.  Since \(S\) is a boson whereas \(D\) is a fermion, the \(DS\) system carries half-integer total angular momentum and cannot annihilate into a purely bosonic final state such as dark photons.  Equivalently, such an annihilation would violate fermion parity. The attractive \(\mathrm{U}(1)_D\) interaction therefore does not erase the primordial dark-sector asymmetry.  Instead, as the dark sector cools, the Coulomb attraction allows \(S\) and \(D\) to form a stable fermionic bound state, \(\mathcal{A}_D\equiv (DS)\), realizing a boson--fermion dark atom after recombination.

Because \(\mathcal{A}_D\) contains one fermionic and one bosonic constituent, $\mathcal{A}_D$ is itself a fermion. In the Coulombic regime, the leading hydrogen-like binding energy and Bohr radius are
\begin{equation}
 E_B=\frac12 \alpha_D^2\mu_{DS},
\qquad
 a_0=\frac{1}{\alpha_D\mu_{DS}} \,,
\end{equation}
where \(\mu_{DS}\) is the reduced mass,
\begin{equation}
\mu_{DS}=\frac{m_D m_S}{m_D+m_S} \,.
\end{equation}
The dark-atom mass is therefore approximately
\begin{equation}
 m_{\mathcal{A}_D}\simeq m_D+m_S-E_B.
\end{equation}
If dark recombination proceeds efficiently, the cosmological dark matter abundance is expected to reside predominantly in $\mathcal{A}_D$, although a residual ionized component consisting of free $D$ and $S$ particles may remain. We will discuss dark recombination in more detail in Sec.~\ref{sec:recombination}. 

Because the atom contains one spin-$1/2$ constituent and one spin-zero constituent, its ground state is a spin-$1/2$ state.  In contrast to a fermion--fermion dark atom, there is no singlet--triplet hyperfine structure in the ground state.  Relativistic and spin--orbit corrections can still split excited states, but the familiar hyperfine phenomenology of dark hydrogen is absent in the present case.

As discussed in the previous section, the symmetric components can efficiently be depleted through annihilations into dark photons, which we assume in the following discussion. The surviving asymmetric populations subsequently recombine to form boson--fermion dark atoms. In the absence of residual ionization and subsequent reannihilation, the observed dark matter density then requires
\begin{align}
Y_\Delta^{\rm obs}
\simeq
7.3\times10^{-13}
\left(\frac{\Omega_{\rm DM}h^2}{0.12}\right)
\left(\frac{m_{\mathcal A_D}}{600~{\rm GeV}}\right)^{-1}\,,
\label{eq:Ydelta-observed}
\end{align}
where we have normalized the DM abundance to the observed
value~\cite{Planck:2018vyg}.

Combining this result with Eq.~\eqref{eq:Ydelta-nonthermal}, we obtain
\begin{align}
\left|
\sum_i B_i^\phi\epsilon_i
\right|
&\simeq
5\times10^{-10} 
\nonumber \\
& \times 
\left(\frac{T_R / m_\phi}{10^{-3}}\right)^{-1}
\left(\frac{\Omega_{\rm DM}h^2}{0.12}\right)
\left(\frac{m_{\mathcal A_D}}{600~{\rm GeV}}\right)^{-1},
\label{eq:required-Bepsilon}
\end{align}
in the absence of washout and entropy dilution. Thus, a dark-atom mass of \(\mathcal{O}(100)~\mathrm{GeV}\) can readily be accommodated for $T_R/m_\phi\sim10^{-3}$ and an effective CP-asymmetric inflaton branching fraction of order $\mathcal{O}( 10^{-10})$--$\mathcal{O}( 10^{-9})$.

Notice that the total number of $D$ and $S$ particles initially produced in the $N_i$ decays is much larger than the net particle--antiparticle asymmetry, since the fractional asymmetry generated in each decay is of order $|\epsilon_i|$. Efficient annihilation of the symmetric components into dark photons is therefore essential for obtaining an asymmetric dark-matter population dominated by the surviving $D$ and $S$ particles.

It is also interesting to note that the value of \(Y_\Delta^{\rm obs}\) in Eq.~\eqref{eq:Ydelta-observed} is not vastly different from the observed baryon asymmetry, \(Y_B\simeq 8.7 \times 10^{-11}\)~\cite{Planck:2018vyg}. In our model, the difference between these two asymmetries can be naturally attributed to the difference between the inflaton branching fraction into the \(N_i\) and that into the right-handed neutrinos, as well as that in the asymmetry parameters. This constitutes an important distinction from conventional asymmetric dark matter scenarios, in which the visible- and dark-sector asymmetries are typically related through transfer interactions or a shared conserved quantum number. In the present framework, inflaton decay instead provides a common origin for the two cosmic abundances: decays into the right-handed neutrinos generate the baryon asymmetry in the visible sector through non-thermal leptogenesis, whereas decays into the \(N_i\) generate the asymmetry that ultimately determines the dark matter abundance.

\section{Thermal decoupling and dark radiation}
\label{sec:dr}

We now discuss the thermal decoupling of the dark sector from the SM plasma and the contribution of the massless dark photon to $\Delta N_{\rm eff}$.

After the symmetric components have annihilated away, the residual asymmetric $D$ and $S$ particles can still act as thermal bridges between the SM plasma and the dark-photon bath. The visible and dark sectors thermally decouple once these residual $D$ and $S$ particles cease to efficiently exchange momentum with the SM thermal bath.

The scalar \(S\) interacts with the SM sector through elastic $Sf\leftrightarrow Sf$ scattering induced by the Higgs portal coupling \(\lambda_{HS}\), where \(f\) denotes a relativistic SM fermion. For $T\ll m_h$, the momentum-transfer rate associated with this process can be estimated as follows. The squared matrix element scales as
\begin{equation}
 |\mathcal{M}_{Sf \to Sf}|^2 \sim \lambda_{HS}^2 \frac{m_f^2 T^2}{m_h^4} \,,
\end{equation}
where \(m_f\) and \(m_h\) denote the masses of the fermion \(f\) and the Higgs boson, respectively. For a more precise treatment, see Ref.~\cite{Binder:2017rgn}. For the heavy scalar \(S\), the corresponding scattering cross section scales as
\begin{equation}
\sigma_{Sf} \sim \lambda_{HS}^2 \frac{m_f^2 T^2}{m_h^4m_S^2} \,.
\end{equation}
The momentum-transfer rate is therefore estimated as
\begin{align}
\gamma_{Sf} \sim n_f \sigma_{Sf} \frac{T}{m_S}\sim \lambda_{HS}^2 \frac{m_f^2 T^6}{m_h^4m_S^3} \,.
\end{align}
Here, the factor $T/m_S$ accounts for the fact that a single collision transfers only a small fraction of the typical momentum of the non-relativistic scalar.

For the \(b\) quark, for example, we obtain
\begin{equation}
\frac{\gamma_{Sb}}{H} \sim 0.2 \left( \frac{\lambda_{HS}}{10^{-4}} \right)^2 \left( \frac{T}{10~\mathrm{GeV}} \right)^4 \left( \frac{m_S}{300~\mathrm{GeV}} \right)^{-3} \,,
\end{equation}
indicating that this interaction can become inefficient at a temperature well above the QCD scale for the representative parameters shown above.

We note that, in the regime of interest, $S$ is non-relativistic and its number density is much smaller than that of the dark photons. Consequently, even if the remaining $S$ particles are still kinetically coupled to the SM plasma, their dilute population may no longer be able to maintain thermal equilibrium between the entire dark-photon bath and the SM sector. The decoupling condition derived above should therefore be regarded as conservative: visible--dark thermal decoupling may occur even for values of $\lambda_{HS}$ larger than those suggested by the kinetic-decoupling condition for an individual $S$ particle.


After visible--dark decoupling, the comoving entropies of the two sectors are separately conserved. If $T_D=T$ at the decoupling temperature \(T_{\rm dec}\), and if all remaining symmetric $D$ and $S$ entropy is eventually transferred to dark photons, entropy conservation gives
\begin{align}
\frac{T_{\gamma_D}}{T_\nu}
=
\left[
\frac{g_{*s}^{\rm SM}(T_{\nu\rm dec})}
{g_{*s}^{\rm SM}(T_{\rm dec})}
\right]^{1/3}
\left[
\frac{g_{*s}^{D}(T_{\rm dec})}{2}
\right]^{1/3},
\label{eq:TD-Tnu}
\end{align}
where \(g_{*s}^{\rm SM}\) and \(g_{*s}^D\) are the effective number of degrees of freedom for entropy density in the SM and in the dark sector, respectively,  $g_{*s}^{\rm SM}(T_{\nu\rm dec})=10.75$, and the factor of $2$ in the denominator accounts for the two helicity states of the massless dark photon. Consequently,
\begin{align}
\Delta N_{\rm eff}^{\gamma_D}
=
\frac{8}{7}
\left(\frac{T_{\gamma_D}}{T_\nu}\right)^4
=
\frac{8}{7}
\left[
\frac{10.75}{g_{*s}^{\rm SM}(T_{\rm dec})}
\frac{g_{*s}^{D}(T_{\rm dec})}{2}
\right]^{4/3}.
\label{eq:DeltaNeff-master}
\end{align}

If visible--dark decoupling occurs at a temperature much lower than \(m_D\) and \(m_S\), \textit{i.e.}, \(T \lesssim 0.1 m_{D,S}\), the entropy carried by the massive dark charged particles is negligible, and the result approaches the irreducible thermal contribution
\begin{align}
\Delta N_{\rm eff}^{\gamma_D}
\simeq
\frac{8}{7}
\left[
\frac{10.75}{g_{*s}^{\rm SM}(T_{\rm dec})}
\right]^{4/3}.
\label{eq:DeltaNeff-floor}
\end{align}
For example, this gives approximately $0.054$, $0.062$, $0.071$, and $0.112$ for $g_{*s}^{\rm SM}=106.75$, $96.25$, $86.25$, and $61.75$, respectively.

A recent combined analysis of primordial abundances, CMB data, and DESI BAO finds \(N_{\rm eff}=2.990\pm0.070\) at 68\% C.L. and quotes the one-sided bound \(\Delta N_{\rm eff}<0.107\) at 95\% C.L.~\cite{Goldstein:2026iuu}. From the above estimates, we find that this bound can be readily satisfied in our scenario provided that visible--dark decoupling occurs at \(T \gtrsim \text{a few}~\mathrm{GeV}\).

It is worth noting that if visible--dark decoupling occurs too early, while $D$ and $S$ are still relativistic, their entropy is subsequently transferred to the dark-photon bath, thereby reheating the dark radiation and enhancing $\Delta N_{\rm eff}$. In this case, the dark-sector entropy degrees of freedom are
\begin{align}
g_{*s}^{D}
=
2+\frac78\times4+2
=
7.5.
\end{align}
Taking $g_{*s}^{\rm SM}=106.75$, we then find
\begin{align}
\Delta N_{\rm eff}^{\gamma_D}
\simeq0.31 \,,
\label{eq:DeltaNeff-early}
\end{align}
which clearly exceeds the above upper bound.

A particularly sharp and testable prediction of our scenario is the presence of dark radiation associated with the massless dark photon.  As discussed above, we predict \(\Delta N_{\rm eff}^{\gamma_D}\simeq0.05\)--0.07 for $g_{*s}^{\rm SM}(T_{\rm dec})\simeq86.25$--$106.75$, placing the model directly within the reach of forthcoming high-precision cosmological observations.  The Simons Observatory is expected to reach \(\sigma(N_{\rm eff})\simeq0.05\)--0.07~\cite{SimonsObservatory:2018koc}, and can therefore begin to probe the predicted range.  CMB-S4-like instruments target a sensitivity of \(\sigma(N_{\rm eff})\simeq0.03\)~\cite{Abazajian:2019eic}, corresponding to an approximately \(2\sigma\) test of our prediction.  A similar sensitivity, \(\sigma(N_{\rm eff})\simeq0.03\), has been forecast for the proposed PICO satellite~\cite{NASAPICO:2019thw}.  Looking further ahead, the proposed CMB-HD survey~\cite{Sehgal:2019ewc, CMB-HD:2022bsz}, when combined with DESI BAO measurements, is forecast to achieve \(\sigma(N_{\rm eff})\simeq0.014\)~\cite{MacInnis:2023vif}, for which the predicted \(\Delta N_{\rm eff}^{\gamma_D}\) would correspond to a \(3.5\)--\(5\sigma\) signal.  Thus, the dark-radiation prediction of the present model is not merely a consistency requirement but constitutes a concrete cosmological target: next-generation CMB observations can begin to probe it, while a CMB-HD-level measurement could either discover the predicted dark radiation with high significance or decisively test the scenario.

\section{Dark recombination}
\label{sec:recombination}
The basic recombination process is
\begin{align}
D+S\leftrightarrow {\cal A}_D+\gamma_D.
\label{eq:recombination-process}
\end{align}
The leading orbital dynamics is Coulombic and is therefore closely related to the standard atomic-dark-matter problem~\cite{Kaplan:2009de, Cyr-Racine:2012tfp, Bansal:2022qbi}.\footnote{Recent work has shown that, for fermion--fermion dark hydrogen, appropriately rescaled SM recombination rates provide a good approximation over a broad range of masses and couplings, including the nearly equal-mass regime~\cite{Barron:2026nks}.} Nevertheless, precise radiative transition rates should ultimately be recomputed for the boson--fermion system.

We define the free ionization fraction by
\begin{align}
x_D
\equiv
\frac{n_D^{\rm free}}
{n_D^{\rm free}+n_{{\cal A}_D}}
=
\frac{n_S^{\rm free}}
{n_S^{\rm free}+n_{{\cal A}_D}}.
\end{align}
Gauge-charge neutrality implies equal free $D$ and $S$ abundances.  Defining
\begin{align}
n_{\rm tot}
\equiv
n_D^{\rm free}+n_{{\cal A}_D}
=
n_S^{\rm free}+n_{{\cal A}_D},
\end{align}
the Saha relation in chemical equilibrium is
\begin{align}
\frac{x_D^2}{1-x_D}
=
\frac{1}{n_{\rm tot}}
\frac{g_Dg_S}{g_{{\cal A}_D}}
\left(\frac{\mu_{DS}T_D}{2\pi}\right)^{3/2}
\exp\left(-\frac{E_B}{T_D}\right),
\label{eq:Saha-general}
\end{align}
where $g_D=2$, $g_S=1$, and $g_{{\cal A}_D}=2$, so that the statistical factor $g_Dg_S/g_{{\cal A}_D}$ is unity.  Here $T_D$ denotes the common matter--radiation temperature of the dark sector as long as kinetic equilibrium between the charged particles and the dark-photon bath is maintained.

As in ordinary hydrogen recombination, the very small particle-photon ratio delays recombination to temperatures well below the binding energy~\cite{Peebles:1968ja}.  Parametrically, the onset of rapid recombination is expected around \(T_D^{\rm rec} \sim 0.02 E_B\)~\cite{Cyr-Racine:2012tfp}.  For illustration, taking \(m_D=m_S=300~{\rm GeV}\) and \(\alpha_D=0.1\) gives \(\mu_{DS}=150~{\rm GeV}\), \(E_B\simeq750~{\rm MeV}\), and thus \(T_D^{\rm rec} \simeq 15~\mathrm{MeV}\). 

An important qualitative difference between the boson-fermion atom and conventional dark hydrogen is that both charged constituents can be heavy.  The Thomson cross section for dark photons scattering on a non-relativistic charged particle $X=D,S$ is
\begin{align}
\sigma_{T,X}
\simeq
\frac{8\pi}{3}\frac{\alpha_D^2}{m_X^2}\,.
\label{eq:dark-Thomson}
\end{align}
The scattering rate of a dark photon prior to recombination is therefore approximately
\begin{align}
\Gamma_{\gamma_D}
\simeq
Y_\Delta s\,
\frac{8\pi\alpha_D^2}{3}
\left(
\frac{1}{m_D^2}
+
\frac{1}{m_S^2}
\right)\,,
\label{eq:dark-photon-opacity}
\end{align}
where \(s\) is the entropy density: 
\begin{equation}
  s(T)=\frac{2\pi^2}{45}\,g_{*s}(T)\,T^3\,,
\end{equation}
with \(g_{*s}(T)\) the effective number of degrees of freedom for entropy density. When $\Gamma_{\gamma_D}\lesssim H$, the mean free path of a dark photon exceeds the Hubble length, so that the dark-photon bath becomes optically thin and begins to free stream.

For $m_D=m_S=m$ and $m_{{\cal A}_D}\simeq2m$, the visible-sector temperature at which this transition occurs is
\begin{align}
T_{\rm fs}
&\simeq
\frac{1.66\cdot135\sqrt{g_*}\,m^2}
{32\pi^3\alpha_D^2g_{*s}M_{\rm Pl}Y_\Delta}
\nonumber\\
&\simeq
70~{\rm MeV}
\left(\frac{m}{300~{\rm GeV}}\right)^3
\left(\frac{\alpha_D}{0.1}\right)^{-2}
\nonumber \\ 
&\quad\times \left(\frac{g_*}{10.75}\right)^{1/2}
\left(\frac{g_{*s}}{10.75}\right)^{-1} \,,
\label{eq:Tgamma-dec-est}
\end{align}
where we have used \(\Omega_{\rm DM}h^2 = 0.12\). 
Notice that $T_{\rm fs}$ in Eq.~\eqref{eq:Tgamma-dec-est} denotes the SM temperature.  If the visible and dark sectors have already thermally decoupled, we need to evaluate the corresponding dark-photon temperature in a similar manner as we have done in Sec.~\ref{sec:dr}.

It is important to distinguish this dark-photon free-streaming condition from kinetic decoupling of the charged DM from the dark-radiation bath~\cite{Cyr-Racine:2012tfp, Bansal:2022qbi}. Even after the dark photons become optically thin, an individual charged particle can continue to experience efficient Compton drag. The corresponding momentum-relaxation rate is parametrically
\begin{align}
\Gamma_{{\rm drag},X}
\simeq
\frac{4\rho_{\gamma_D}}{3m_X}\,
\sigma_{T,X}
=
\frac{32\pi}{9}
\frac{\alpha_D^2\rho_{\gamma_D}}{m_X^3}\,,
\end{align}
where \(\rho_{\gamma_D} = \pi^2 T_D^4/15\) denotes the dark-photon energy density. For the benchmark above, this rate remains much larger than the Hubble expansion rate at $T_D\sim T_D^{\rm rec}$.  Thus the charged $D$ and $S$ populations are expected to remain approximately kinetically coupled to the dark-photon bath until dark recombination substantially reduces the free-charge abundance.  In this parameter region, recombination itself, rather than the onset of dark-photon free streaming, is therefore likely to terminate the efficient matter--radiation coupling.  A quantitative treatment of this transition requires following the ionization fraction and the matter and dark-photon temperatures simultaneously~\cite{Cyr-Racine:2012tfp,Bansal:2022qbi}.

To estimate the residual ionized fraction, we use the approximate analytic result obtained for hydrogen-like atomic dark matter~\cite{Cyr-Racine:2012tfp},
\begin{align}
x_D^\infty
\sim
2\times10^{-16}
\frac{\xi}{\alpha_D^6}
\left(\frac{m_{{\cal A}_D}}{{\rm GeV}}\right)
\left(\frac{E_B}{{\rm keV}}\right),
\label{eq:residual-ionization-est}
\end{align}
where \(\xi\equiv T_D/T\) denotes the dark-sector-to-SM temperature ratio.  This expression should be regarded only as an order-of-magnitude estimate~\cite{Cyr-Racine:2012tfp}.  For \(m_D=m_S=m\), we have \(m_{{\cal A}_D}\simeq2m\) and \(E_B=\alpha_D^2m/4\), and Eq.~\eqref{eq:residual-ionization-est} becomes
\begin{align}
x_D^\infty
\sim
4.5\times10^{-2}
\left(\frac{\xi}{0.5}\right)
\left(\frac{\alpha_D}{0.1}\right)^{-4}
\left(\frac{m}{300~{\rm GeV}}\right)^2.
\label{eq:residual-ionization-benchmark}
\end{align}
Thus, even this simple estimate indicates that the residual ionized fraction is already reduced to the few-percent level for our representative benchmark.  The result is particularly sensitive to the dark gauge coupling, with \(x_D^\infty\propto\alpha_D^{-4}\) after using \(E_B\propto\alpha_D^2\).

Equation~\eqref{eq:residual-ionization-est} assumes a common temperature for the charged particles and the dark-photon bath during recombination. In parameter regions where the charged particles thermally decouple from the dark radiation before recombination and consequently become colder than the dark-photon bath, low-velocity radiative capture is enhanced,
and the resulting residual ionized fraction can be smaller than the estimate above~\cite{Cyr-Racine:2012tfp, Bansal:2022qbi}.  A dedicated calculation of boson--fermion recombination, including the coupled thermal evolution of the charged particles and dark photons, would be required for a quantitatively reliable prediction~\cite{Cyr-Racine:2012tfp,Bansal:2022qbi,Barron:2026nks}.

We emphasize that efficient recombination is not guaranteed throughout the full parameter space of the model.  In particular, because of the strong dependence on \(\alpha_D\), a sizable or even dominant ionized component may survive for smaller dark gauge couplings.  In such regions, the coupled evolution of the ionization fraction, the charged-particle temperature, and the dark-photon temperature must be treated explicitly, together with the associated cosmological and astrophysical constraints.  A detailed analysis of this partially ionized regime is beyond the scope of the present work.  In the following analysis, we therefore assume that dark recombination is sufficiently efficient that the residual ionized fraction is phenomenologically negligible and that the dark-matter abundance is dominated by the neutral boson--fermion atoms ${\cal A}_D$, as can be achieved for sufficiently large \(\alpha_D\).

In the rest of this section, however, we briefly discuss phenomenological consequences which a non-negligible residual ionized component may have, just for convenience. First, the free $D$ and $S$ particles remain coupled to the massless dark photon through long-range gauge interactions.  In the early Universe, scattering with the dark-radiation bath can therefore delay the growth of density perturbations and generate dark acoustic oscillations or a smooth suppression of the matter power spectrum on sufficiently small scales~\cite{Cyr-Racine:2012tfp, Bansal:2022qbi, Barron:2026nks}.  The strength of this effect depends not only on the residual ionized fraction but also on the dark-radiation temperature and on the epoch at which the charged component kinetically decouples.

Second, residual dark charges undergo long-range Coulomb scattering inside DM halos.  The corresponding momentum-transfer cross section has the characteristic low-velocity behavior~\cite{Feng:2009mn, Agrawal:2016quu}
\begin{align}
\sigma_T^{ij}
\sim
\frac{4\pi\alpha_D^2}
{\mu_{ij}^2v^4}
\ln\Lambda_D,
\label{eq:dark-Coulomb-transfer}
\end{align}
where \(i,j=D,S\) label the species of the free dark charged particles participating in the scattering, \(\mu_{ij}=m_i m_j/(m_i+m_j)\) is their reduced mass, \(v\) denotes their relative velocity, and \(\ln\Lambda_D\) is the dark-sector Coulomb logarithm.  In a virialized halo, \(v\) is typically of the order of the characteristic velocity dispersion of the halo. The strong $v^{-4}$ enhancement makes this interaction particularly relevant in low-velocity systems such as dwarf galaxies.  Efficient Coulomb scattering can isotropize the charged component and induce heat transport within halos, potentially modifying halo shapes, density profiles, and their gravothermal evolution~\cite{Feng:2009mn,Cyr-Racine:2012tfp}.  For a partially ionized dark-matter population, however, the impact on the total halo cannot be inferred from the ionized fraction alone, since it also depends on the relative spatial distributions of the ionized and neutral components and on the efficiency of momentum and energy exchange between them.  A quantitative constraint therefore requires a simultaneous treatment of ion--ion, ion--atom, and atom--atom scattering, and in the regime with an intermediate ionized fraction a dedicated halo analysis may be necessary~\cite{Kaplan:2011yj,Cyr-Racine:2012tfp, Petraki:2014uza}. 

In addition, the residual plasma can undergo dissipative processes such as dark bremsstrahlung and radiative recombination~\cite{Fan:2013tia, Rosenberg:2017qia}.
These processes may lead to cooling and further evolution of the ionized component inside halos~\cite{Rosenberg:2017qia,Ghalsasi:2017jna}. Conversely, continued recombination in sufficiently cool halos can further reduce the free charged fraction. The resulting astrophysical phenomenology is consequently sensitive to the detailed ionization and thermal history and cannot, in general, be characterized by a universal upper bound on $x_D^\infty$~\cite{Ghalsasi:2017jna, Roy:2023zar}. Radiative bound-state formation can also provide indirect-detection signals in asymmetric dark matter models when the emitted mediator subsequently decays into SM particles~\cite{Pearce:2013ola, Pearce:2015zca}. In the present scenario, however, the emitted dark photons are massless and their kinetic mixing with hypercharge is assumed to be negligible, so this mechanism does not yield the visible signals considered in those studies.

\section{Dark Atom--Antiatom oscillations}
\label{sec:oscillation}

After dark recombination, the $X$-violating operators in Eq.~\eqref{eq:effective-washout} induce mixing between ${\cal A}_D$ and its antiparticle $\overline{\cal A}_D$. Oscillations can regenerate antiatoms and allow annihilation after the initial symmetric components have been depleted~\cite{Buckley:2011ye, Cirelli:2011ac, Tulin:2012re, Okada:2012rm}. We examine the onset of oscillations and estimate the resulting change in the dark matter abundance. For simplicity, we focus on equal constituent masses, \(m_D=m_S=m\), so that \(\mu_{DS}=m/2\) and \(m_{{\cal A}_D}\simeq2m\).

Defining \(C_{\rm mix}\equiv C_g+C_{\widetilde g}^{*}\), the leading non-relativistic interaction responsible for atom--antiatom conversion is described by the contact potential
\begin{align}
V_{\overline{\cal A}_D{\cal A}_D}(\boldsymbol r)
=
\frac{C_{\rm mix}}{2m_S}
\delta^{(3)}(\boldsymbol r)\,.
\end{align}
The factor \(1/m_S\) arises from the non-relativistic normalization of the scalar field. The off-diagonal matrix element between ground-state atoms and antiatoms with the same spin state is
\begin{align}
\delta_A
&\equiv
\langle\overline{\cal A}_D|H|{\cal A}_D\rangle
=
\frac{C_{\rm mix}}{2m_S}|\psi_{1s}(0)|^2\,,
\label{eq:osc-matrix-element}
\end{align}
where
\begin{align}
\psi_{1s}(r)
=
\frac{e^{-r/a_0}}{\sqrt{\pi a_0^3}}\,,
\qquad
|\psi_{1s}(0)|^2
=
\frac{(\alpha_D\mu_{DS})^3}{\pi}\,.
\end{align}
For equal constituent masses, this gives
\begin{align}
\delta_A
=
\frac{C_{\rm mix}\alpha_D^3m^2}{16\pi}\,.
\end{align}
The mass eigenstates are separated by \(2|\delta_A|\), and an initially pure atom state converts into an antiatom in vacuum with probability
\begin{align}
P_{{\cal A}_D\to\overline{\cal A}_D}(t)
=
\sin^2(|\delta_A|t)\,.
\end{align}

For the representative parameters used in Sec.~\ref{sec:recombination},
\begin{align}
|\delta_A|
&\simeq
1.8\times10^{-12}\ {\rm GeV}
\left(
\frac{|C_{\rm mix}|}{10^{-12}\ {\rm GeV}^{-1}}
\right)
\nonumber\\
&\quad\times
\left(\frac{\alpha_D}{0.1}\right)^3
\left(\frac{m}{300\ {\rm GeV}}\right)^2.
\label{eq:osc-delta-benchmark}
\end{align}
Taking \(\xi=T_D/T=0.5\), the dark recombination temperature \(T_D^{\rm rec}\sim15\ {\rm MeV}\) corresponds to \(T^{\rm rec}\sim30\ {\rm MeV}\) and \(H_{\rm rec}\simeq4\times10^{-22}\ {\rm GeV}\). Thus, unless \(C_{\mathrm{mix}}\) is strongly suppressed, the vacuum oscillation rate is much larger than the expansion rate when dark atoms form; for this benchmark, delaying vacuum oscillations beyond recombination requires \(|C_{\rm mix}|\lesssim2\times10^{-22}\ {\rm GeV}^{-1}\). 

The atom--antiatom mixing can equivalently be described in terms of Majorana mass eigenstates. The $X$-violating interactions split the approximately Dirac dark atom into two nearly degenerate Majorana fermions, denoted by $\chi_1$ and $\chi_2$, with masses
\begin{align}
m_1 &= m_{{\cal A}_D}-|\delta_A|\,,
&
m_2 &= m_{{\cal A}_D}+|\delta_A|\,.
\end{align}
Here, common mass corrections are understood to be included in $m_{{\cal A}_D}$. The physical mass splitting is therefore
\begin{align}
\Delta m \equiv m_2-m_1=2|\delta_A|\,.
\label{eq:Majorana-atom-splitting}
\end{align}
An initially produced dark atom is a coherent superposition of these two mass eigenstates. Their relative phase evolves at a frequency $\Delta m$, reproducing the atom--antiatom oscillation probability given above. The oscillation and Majorana descriptions thus represent the same physics in different bases.

Oscillations alone preserve the total number of atoms and antiatoms. A reduction in the dark matter abundance requires annihilation involving the regenerated antiatoms. To estimate this effect, we parameterize the averaged annihilation cross section of an atom--antiatom pair as
\begin{align}
\langle\sigma_{\rm ann}v_{\rm rel}\rangle
=
c_A\frac{\pi\alpha_D^2}{m^2}\,,
\label{eq:osc-annihilation-coefficient}
\end{align}
where \(c_A\) is a dimensionless coefficient that we take to be constant. Numerically,
\begin{align}
\langle\sigma_{\rm ann}v_{\rm rel}\rangle
&\simeq
4.1\times10^{-24}~{\rm cm^3\,s^{-1}} \nonumber \\ 
& \times c_A
\left(\frac{\alpha_D}{0.1}\right)^2
\left(\frac{m}{300\ {\rm GeV}}\right)^{-2}.
\end{align}
The coefficient $c_A$ normalizes the effective annihilation
rate to the constituent annihilation scale and need not be
smaller than unity. We take $c_A$ to be constant and of order
unity as a benchmark appropriate to direct annihilation
without large atomic-collision effects. Short-distance
correlations, rearrangement reactions~\cite{Qiu:2023bbp}, and near-threshold
resonances can substantially enhance or suppress the rate
and may introduce a nontrivial velocity dependence.

For the abundance estimate, we assume that oscillations establish an approximately equal, incoherent population,\footnote{Rapid coherent oscillations alone do not generally justify this classical population description; a treatment retaining coherence would require density-matrix equations~\cite{Tulin:2012re}. We use the equal-population limit to obtain a simple estimate of reannihilation.}
\begin{align}
n_{{\cal A}_D}
\simeq
n_{\overline{\cal A}_D}
\simeq
\frac{n_A}{2}\,,
\qquad
Y_A\equiv\frac{n_A}{s}\,,
\end{align}
where \(n_A\) is the total atom-plus-antiatom number density. 

Neglecting inverse annihilation and assuming conserved SM comoving entropy, the total yield obeys
\begin{align}
\frac{dY_A}{dt}
=
-\frac{s\langle\sigma_{\rm ann}v_{\rm rel}\rangle}{2}Y_A^2\,.
\label{eq:osc-yield-evolution}
\end{align}
The factor \(1/2\) follows because each annihilation removes two particles, while the pair density is \(n_A^2/4\). If \(t_{\rm on}\) denotes the onset of the effective annihilating population, integration gives
\begin{align}
\frac{Y_A(\infty)}{Y_A(t_{\rm on})}
&=
\frac{1}{1+\tau_A}\,,
\end{align}
where 
\begin{align}
  \tau_A
\equiv
\frac{Y_A(t_{\rm on})\langle\sigma_{\rm ann}v_{\rm rel}\rangle}{2}
\int_{t_{\rm on}}^\infty s(t)\,dt\,.
\label{eq:osc-optical-depth}
\end{align}
For constant \(\langle\sigma_{\rm ann}v_{\rm rel}\rangle\), the integral is dominated by early times. Approximating the relativistic degrees of freedom as constant during radiation domination, we obtain
\begin{align}
\tau_A
&\simeq
\left.\frac{n_A\langle\sigma_{\rm ann}v_{\rm rel}\rangle}{2H}\right|_{\rm on}
\nonumber\\
&=
0.132\frac{g_{*s}}{\sqrt{g_*}}
M_{\rm Pl}T_{\rm on}
Y_A(t_{\rm on}) \cdot 
c_A\frac{\pi\alpha_D^2}{m^2}\,.
\label{eq:osc-survival}
\end{align}
The fraction removed by reannihilation is therefore
\begin{align}
f_{\rm dep}
\equiv
1-\frac{Y_A(\infty)}{Y_A(t_{\rm on})}
=
\frac{\tau_A}{1+\tau_A}\,.
\end{align}

We first take \(t_{\rm on}\simeq t_{\rm rec}\) and \(Y_A(t_{\rm on})\simeq Y_\Delta\). For the reference parameters,
\begin{align}
\tau_A
&\simeq
4.0\times10^{-2}c_A
\left(\frac{Y_\Delta}{7.3\times10^{-13}}\right)
\left(\frac{T_{\rm on}}{30\ {\rm MeV}}\right)
\nonumber\\
&\quad\times
\left(\frac{\alpha_D}{0.1}\right)^2
\left(\frac{m}{300\ {\rm GeV}}\right)^{-2}
\nonumber\\
&\quad\times
\left(\frac{g_{*s}}{10.75}\right)
\left(\frac{g_*}{10.75}\right)^{-1/2}.
\label{eq:osc-depletion-benchmark}
\end{align}
Thus, \(c_A=1\) gives \(f_{\rm dep}\simeq0.039\): oscillations regenerate antiatoms, but subsequent annihilation removes only a small fraction of the total abundance. Requiring less than ten percent depletion gives \(\tau_A<1/9\), or approximately \(c_A\lesssim2.8\) for this benchmark. 

The parameter dependence becomes particularly transparent when oscillations are effective immediately after recombination. Using
\begin{align}
T_{\rm on}
\simeq T^{\rm rec}
\simeq
\frac{0.02E_B}{\xi}
=
\frac{0.005\alpha_D^2m}{\xi}\,,
\end{align}
together with the reference asymmetric yield in Eq.~\eqref{eq:Ydelta-observed}, we find
\begin{align}
\tau_A
&\simeq
4.0\times10^{-2}c_A
\left(\frac{\Omega_{\rm DM}h^2}{0.12}\right)
\left(\frac{\xi}{0.5}\right)^{-1}
\nonumber\\
&\quad\times
\left(\frac{\alpha_D}{0.1}\right)^4
\left(\frac{m}{300\ {\rm GeV}}\right)^{-2}
\nonumber\\
&\quad\times
\left(\frac{g_{*s}}{10.75}\right)
\left(\frac{g_*}{10.75}\right)^{-1/2}.
\label{eq:osc-depletion-scaling}
\end{align}
The scaling \(\tau_A\propto c_A\alpha_D^4/m^2\) reflects both the larger annihilation coefficient and the earlier recombination at stronger coupling. Consequently, increasing \(\alpha_D\) to improve recombination can also increase reannihilation.

For completeness, suppose that the same constant coefficient applies in present-day halos. For equal atom and antiatom densities, the annihilation rate per atom is
\begin{align}
\Gamma_{\rm ann}
=
\frac{\rho_{\rm DM}}{2m_{{\cal A}_D}} \langle\sigma_{\rm ann}v_{\rm rel}\rangle
\simeq
\frac{\rho_{\rm DM}}{4m} \langle\sigma_{\rm ann}v_{\rm rel}\rangle\,.
\end{align}
At a local density \(\rho_{\rm DM}=0.3\ {\rm GeV\,cm^{-3}}\), this gives
\begin{align}
\Gamma_{\rm ann}t_0
&\simeq
4.5\times10^{-10}c_A
\left(
\frac{\rho_{\rm DM}}{0.3\ {\rm GeV\,cm^{-3}}}
\right)
\nonumber\\
&\quad\times
\left(\frac{\alpha_D}{0.1}\right)^2
\left(\frac{m}{300\ {\rm GeV}}\right)^{-3}
\left(\frac{t_0}{4.4\times10^{17}\ {\rm s}}\right).
\end{align}
Ordinary halo mass loss is therefore negligible for \(\mathcal{O}(1)\) \(c_A\). In the scenario considered here, annihilation into dark photons does not produce a visible indirect-detection signal. Such signals could arise in extensions of the model in which the dark photons can decay into SM particles.

The Majorana description described above reveals a radiative transition between the two states. Although the dark atom is electrically neutral, it possesses a dark magnetic dipole moment inherited from its fermionic constituent. In the $1s$ ground state, the orbital angular momentum vanishes and the scalar constituent carries no intrinsic spin. Consequently, the leading magnetic interaction is
\begin{align}
H_{\rm mag}
=-\mu_{\cal A}\,
\boldsymbol{\sigma}\cdot\boldsymbol{B}_D\,,
\qquad
\mu_{\cal A}\simeq\frac{g_D}{2m_D}\,,
\label{eq:atomic-dark-magnetic-moment}
\end{align}
where $\boldsymbol{B}_D$ is the dark magnetic field and $\boldsymbol{\sigma}$ acts on the atomic spin. We neglect radiative and relativistic bound-state corrections to the magnetic moment. The magnetic moments of an atom and an antiatom have opposite signs. Upon transforming to the Majorana mass basis, this interaction becomes off-diagonal, giving a transition magnetic moment
\begin{align}
|\mu_{12}|\simeq\frac{g_D}{2m_D}\,.
\label{eq:Majorana-transition-moment}
\end{align}
This is consistent with the absence of diagonal magnetic dipole moments for Majorana fermions. 

The heavier state can consequently undergo the magnetic-dipole transition \(\chi_2\to\chi_1+\gamma_D\). Since the dark photon is massless, the decay is kinematically allowed for any positive $\Delta m$ in vacuum. With the transition moment normalized as in Eq.~\eqref{eq:Majorana-transition-moment}, the vacuum decay width is
\begin{align}
\Gamma_2
&=
\frac{|\mu_{12}|^2}{8\pi}
\frac{(m_2^2-m_1^2)^3}{m_2^3}
\nonumber\\
&\simeq
\frac{|\mu_{12}|^2}{\pi}(\Delta m)^3
\simeq
\frac{\alpha_D}{m_D^2}(\Delta m)^3\,.
\label{eq:Majorana-atom-decay-width}
\end{align}
For equal constituent masses, $m_D=m_S=m$, the corresponding vacuum lifetime is estimated using Eq.~\eqref{eq:osc-delta-benchmark} as 
\begin{align}
\tau_2
&\simeq
1.3\times10^{16}\ {\rm s}
\left(\frac{|C_{\rm mix}|}
{10^{-12}\ {\rm GeV}^{-1}}\right)^{-3}
\nonumber\\
&\quad\times
\left(\frac{\alpha_D}{0.1}\right)^{-10}
\left(\frac{m}{300\ {\rm GeV}}\right)^{-4}.
\label{eq:Majorana-lifetime-benchmark}
\end{align}
Thus, the reference parameters give a lifetime of $\simeq 4\times10^8$ years. The radiative decay is negligible during dark recombination but can substantially deplete the heavier Majorana population over cosmological times.

For the reference parameters, the present dark-photon temperature is well below the splitting and the vacuum lifetime is shorter than the age of the Universe. In the absence of efficient late-time re-excitation, the present dark matter is therefore expected to reside predominantly in $\chi_1$. This decay preserves the total number of dark atoms and releases only the small mass difference, with $\Delta m/m_{{\cal A}_D}\simeq6\times10^{-15}$ at the benchmark point. It consequently has a negligible direct effect on the dark matter energy density and on the dark-radiation abundance.

In summary, the mixing operators can induce rapid atom--antiatom
oscillations as soon as dark atoms form. Under the assumed
incoherent, equal-population approximation, reannihilation
depletes the total abundance by only a few percent for
$c_A=1$ at the reference masses and coupling. The late-time
population can therefore contain comparable numbers of atoms
and antiatoms while its total abundance remains determined
mainly by the primordial asymmetry. This result relies on
the benchmark assumption of a constant, order-unity $c_A$.
Atomic-collision effects may enhance or suppress the effective
annihilation rate and introduce velocity dependence, so relic
survival is not guaranteed throughout the parameter space.
If reannihilation is more efficient than assumed here,
reproducing the observed DM density would require a
correspondingly larger primordial asymmetry to compensate
for the additional depletion. A microscopic calculation
of atom--antiatom reactions is therefore required to determine
the magnitude and velocity dependence of the rate and to
establish the relation between the primordial asymmetry
and the present-day DM abundance.

\section{Discussion}
\label{sec:discussion}

In this work, we have proposed a simple realization of boson--fermion atomic dark matter. The dark-sector asymmetry is generated non-thermally through the CP-violating decays of heavy Majorana fermions produced by inflaton decay, in close analogy with non-thermal leptogenesis in the visible sector. An important feature of this framework is that the asymmetry is shared between a scalar and a fermion carrying opposite dark charges. Their symmetric components can efficiently annihilate into dark photons, whereas the remaining asymmetric populations cannot annihilate with each other and instead recombine into neutral boson--fermion dark atoms.

This setup provides an interesting connection between asymmetric dark matter and atomic dark matter. The dark matter abundance is initially determined by the primordial asymmetry, while the same interactions responsible for its generation induce atom--antiatom mixing after recombination, splitting the dark atom into two nearly degenerate Majorana states. The present-day population can therefore be symmetric despite its asymmetric origin, either as a mixture of atoms and antiatoms or, after radiative decay of the heavier Majorana state, as a population dominated by the lighter state. Annihilation is consequently possible, although its rate depends on the late-time population and the relevant atomic reaction amplitudes. Under the assumptions adopted in our benchmark estimate, reannihilation changes the relic abundance only moderately; a more precise treatment requires a microscopic calculation of atomic scattering and annihilation. In the scenario considered here, annihilation into massless dark photons does not produce a visible indirect-detection signal. Such signals could nevertheless arise in extensions of the model in which the dark photons acquire a mass and can decay into SM particles via kinetic mixing.

The massless dark photon provides another characteristic prediction. For the thermal history considered in this work, the dark radiation contribution is typically \(\Delta N_{\rm eff}\simeq0.05\)--0.07. This range can be probed by future precision CMB observations and therefore offers a particularly direct test of the scenario. Additional cosmological and astrophysical signatures may arise from dark matter self-interactions. For example, a lighter fermionic constituent, \(m_D\ll m_S\), reduces the reduced mass \(\mu_{DS}\simeq m_D\) and increases the atomic size, \(a_0=(\alpha_D\mu_{DS})^{-1}\), at fixed \(\alpha_D\). The resulting atom--atom scattering may become sufficiently strong~\cite{Qiu:2023bbp} to affect the density profiles and shapes of dark matter halos, providing further opportunities to test the model with astrophysical observations. The magnitude and velocity dependence of these interactions require a dedicated scattering calculation. A smaller reduced mass also lowers the binding energy and shifts dark recombination to lower temperatures. Its effect on the residual ionized fraction, however, depends on both the radiative capture rates and the thermal history, and cannot be inferred from the binding energy alone. Any surviving ionized component may lead to additional cosmological and astrophysical effects. A dedicated calculation of boson--fermion recombination, atomic scattering, and the evolution of the partially ionized dark plasma would therefore help identify the observable signatures and viable parameter space of the model.

Although we have focused on an asymmetric production mechanism, boson--fermion atomic dark matter could in principle also be realized with a symmetric thermal population. Such a realization is, however, expected to be more restrictive. In the symmetric case, the relic abundances of the fermion $D$ and the scalar $S$ are not generically equal, so that efficient recombination into neutral dark atoms does not automatically ensure a sufficiently small residual ionized fraction; an excess of either species may remain unbound. Moreover, reducing the ionized fraction generally favors a relatively large $\mathrm{U}(1)_D$ gauge coupling, which enhances annihilation of the symmetric components and therefore tends to suppress the total dark-matter relic abundance. This is qualitatively different from the asymmetric case, where the final abundance is primarily set by the primordial asymmetry and can remain sizable even when annihilation is very efficient. One may attempt to compensate for the enhanced annihilation by increasing the masses of $D$ and $S$, but larger constituent masses can in turn make recombination less efficient and increase the residual ionized fraction. These competing requirements suggest that realizing boson--fermion atomic dark matter through a symmetric production mechanism may require a more restricted region of parameter space than in the asymmetric scenario considered here. Determining whether a viable symmetric realization exists nevertheless requires a dedicated calculation of the relic abundances and atomic recombination history, and a quantitative comparison with the asymmetric case would be an interesting subject for future study.

More generally, the present framework illustrates how the visible- and dark-matter abundances can have a common cosmological origin without requiring an interaction that directly transfers an asymmetry between the two sectors. Inflaton decays can independently seed the baryon and dark-sector asymmetries, while the subsequent dark-sector dynamics gives rise to atomic dark matter, dark radiation, and particle--antiparticle oscillations. It is intriguing in this regard that the mechanisms responsible for generating the two asymmetries are structurally very similar: in both sectors, the CP asymmetry arises from the interference between tree-level and one-loop amplitudes in the decays of heavy Majorana fermions. In the absence of large hierarchies in the relevant couplings and mass parameters, it is therefore natural to expect the CP-asymmetry parameters in the visible and dark sectors to be comparable in order of magnitude. If, in addition, the corresponding inflaton branching fractions are not strongly hierarchical, the resulting visible- and dark-sector number asymmetries may also be of comparable size, up to sphaleron-conversion, washout, and other efficiency factors. In such a case, the observed ratio of the dark-matter and baryon energy densities points to a characteristic dark-atom mass in the \(\mathcal{O}(10)\,\mathrm{GeV}\) range, as is familiar in asymmetric-dark-matter scenarios~\cite{An:2009vq,Falkowski:2011xh,Josse-Michaux:2011sjn,Narendra:2018vfw}. It would be particularly interesting to investigate whether the present framework naturally realizes such a mass scale and whether this qualitative observation can be sharpened into a more quantitative relation between the visible- and dark-sector abundances. This mass range lies below the representative benchmark masses adopted in this work, which were chosen mainly for illustration, and exploring the lighter-mass regime would require a dedicated analysis of symmetric-component depletion, dark recombination, and the subsequent cosmological evolution. More broadly, these features highlight the possibility that, just as the visible sector exhibits a rich structure of particles, forces, and bound states, the dark sector may possess comparably rich dynamics rather than consisting of a single stable particle. The framework presented here provides a particularly simple and concrete realization of such a dark sector, arising from only a small number of new fields and interactions. Further exploration of its cosmological and astrophysical signatures may therefore provide a window into the possible structure of the dark world.

\begin{acknowledgments}
The work of N.N. was supported in part by the Grant-in-Aid for Scientific Research C (No. 25K07314). T.~T.~Y. is supported by MEXT Grant No.~24H02244 and the World Premier International Research Center Initiative (WPI), MEXT, Japan (Kavli IPMU).

\end{acknowledgments}

\bibliographystyle{name}
\bibliography{references}

@article{Krauss:1988zc,
    author = "Krauss, Lawrence M. and Wilczek, Frank",
    title = "{Discrete Gauge Symmetry in Continuum Theories}",
    reportNumber = "YCTP-P26-88, NSF-ITP-88-187",
    doi = "10.1103/PhysRevLett.62.1221",
    journal = "Phys. Rev. Lett.",
    volume = "62",
    pages = "1221",
    year = "1989"
}

@article{Ibanez:1991hv,
    author = "Ibanez, Luis E. and Ross, Graham G.",
    title = "{Discrete gauge symmetry anomalies}",
    doi = "10.1016/0370-2693(91)91614-2",
    journal = "Phys. Lett. B",
    volume = "260",
    pages = "291--295",
    year = "1991"
}

@article{Ibanez:1991pr,
    author = "Ibanez, Luis E. and Ross, Graham G.",
    title = "{Discrete gauge symmetries and the origin of baryon and lepton number conservation in supersymmetric versions of the standard model}",
    reportNumber = "CERN-TH-6111-91",
    doi = "10.1016/0550-3213(92)90195-H",
    journal = "Nucl. Phys. B",
    volume = "368",
    pages = "3--37",
    year = "1992"
}

@article{Martin:1992mq,
    author = "Martin, Stephen P.",
    title = "{Some simple criteria for gauged R-parity}",
    eprint = "hep-ph/9207218",
    archivePrefix = "arXiv",
    reportNumber = "UFIFT-HEP-92-22",
    doi = "10.1103/PhysRevD.46.R2769",
    journal = "Phys. Rev. D",
    volume = "46",
    pages = "R2769--R2772",
    year = "1992"
}

@article{Holdom:1985ag,
    author = "Holdom, Bob",
    title = "{Two U(1)'s and Epsilon Charge Shifts}",
    reportNumber = "UTPT-85-30",
    doi = "10.1016/0370-2693(86)91377-8",
    journal = "Phys. Lett. B",
    volume = "166",
    pages = "196--198",
    year = "1986"
}

@article{Fukugita:1986hr,
    author = "Fukugita, M. and Yanagida, T.",
    title = "{Baryogenesis Without Grand Unification}",
    reportNumber = "RIFP-641",
    doi = "10.1016/0370-2693(86)91126-3",
    journal = "Phys. Lett. B",
    volume = "174",
    pages = "45--47",
    year = "1986"
}

@article{Falkowski:2011xh,
    author = "Falkowski, Adam and Ruderman, Joshua T. and Volansky, Tomer",
    title = "{Asymmetric Dark Matter from Leptogenesis}",
    eprint = "1101.4936",
    archivePrefix = "arXiv",
    primaryClass = "hep-ph",
    reportNumber = "LPT-ORSAY-11-09",
    doi = "10.1007/JHEP05(2011)106",
    journal = "JHEP",
    volume = "05",
    pages = "106",
    year = "2011"
}

@article{Josse-Michaux:2011sjn,
    author = "Josse-Michaux, Francois-Xavier and Molinaro, Emiliano",
    title = "{A Common Framework for Dark Matter, Leptogenesis and Neutrino Masses}",
    eprint = "1108.0482",
    archivePrefix = "arXiv",
    primaryClass = "hep-ph",
    doi = "10.1103/PhysRevD.84.125021",
    journal = "Phys. Rev. D",
    volume = "84",
    pages = "125021",
    year = "2011"
}

@article{Narendra:2018vfw,
    author = "Narendra, Nimmala and Patra, Sudhanwa and Sahu, Narendra and Shil, Sujay",
    title = "{Baryogenesis via Leptogenesis from Asymmetric Dark Matter and radiatively generated Neutrino mass}",
    eprint = "1805.04860",
    archivePrefix = "arXiv",
    primaryClass = "hep-ph",
    reportNumber = "IP-BBSR-2018-7",
    doi = "10.1103/PhysRevD.98.095016",
    journal = "Phys. Rev. D",
    volume = "98",
    number = "9",
    pages = "095016",
    year = "2018"
}

@article{Asaka:1999yd,
    author = "Asaka, T. and Hamaguchi, Koichi and Kawasaki, M. and Yanagida, T.",
    title = "{Leptogenesis in inflaton decay}",
    eprint = "hep-ph/9906366",
    archivePrefix = "arXiv",
    reportNumber = "UT-853, RESCEU-15-99",
    doi = "10.1016/S0370-2693(99)01020-5",
    journal = "Phys. Lett. B",
    volume = "464",
    pages = "12--18",
    year = "1999"
}

@article{Lazarides:1990huy,
    author = "Lazarides, George and Shafi, Q.",
    title = "{Origin of matter in the inflationary cosmology}",
    reportNumber = "BA-90-78",
    doi = "10.1016/0370-2693(91)91090-I",
    journal = "Phys. Lett. B",
    volume = "258",
    pages = "305--309",
    year = "1991"
}

@article{Kumekawa:1994gx,
    author = "Kumekawa, Kazuya and Moroi, Takeo and Yanagida, Tsutomu",
    title = "{Flat potential for inflaton with a discrete R invariance in supergravity}",
    eprint = "hep-ph/9405337",
    archivePrefix = "arXiv",
    reportNumber = "TU-458",
    doi = "10.1143/PTP.92.437",
    journal = "Prog. Theor. Phys.",
    volume = "92",
    pages = "437--448",
    year = "1994"
}

@article{Asaka:1999jb,
    author = "Asaka, T. and Hamaguchi, Koichi and Kawasaki, M. and Yanagida, T.",
    title = "{Leptogenesis in inflationary universe}",
    eprint = "hep-ph/9907559",
    archivePrefix = "arXiv",
    reportNumber = "UT-855, RESCEU-28-99",
    doi = "10.1103/PhysRevD.61.083512",
    journal = "Phys. Rev. D",
    volume = "61",
    pages = "083512",
    year = "2000"
}

@article{Chung:1998rq,
    author = "Chung, Daniel J. H. and Kolb, Edward W. and Riotto, Antonio",
    title = "{Production of massive particles during reheating}",
    eprint = "hep-ph/9809453",
    archivePrefix = "arXiv",
    reportNumber = "FERMILAB-PUB-98-217-A, CERN-TH-98-227",
    doi = "10.1103/PhysRevD.60.063504",
    journal = "Phys. Rev. D",
    volume = "60",
    pages = "063504",
    year = "1999"
}

@article{Giudice:2000ex,
    author = "Giudice, Gian Francesco and Kolb, Edward W. and Riotto, Antonio",
    title = "{Largest temperature of the radiation era and its cosmological implications}",
    eprint = "hep-ph/0005123",
    archivePrefix = "arXiv",
    reportNumber = "SNS-PH-00-05, FERMILAB-PUB-00-075-A, CERN-TH-2000-107",
    doi = "10.1103/PhysRevD.64.023508",
    journal = "Phys. Rev. D",
    volume = "64",
    pages = "023508",
    year = "2001"
}

@article{Kaplan:2009de,
    author = "Kaplan, David E. and Krnjaic, Gordan Z. and Rehermann, Keith R. and Wells, Christopher M.",
    title = "{Atomic Dark Matter}",
    eprint = "0909.0753",
    archivePrefix = "arXiv",
    primaryClass = "hep-ph",
    doi = "10.1088/1475-7516/2010/05/021",
    journal = "JCAP",
    volume = "05",
    pages = "021",
    year = "2010"
}

@article{Kaplan:2011yj,
    author = "Kaplan, David E. and Krnjaic, Gordan Z. and Rehermann, Keith R. and Wells, Christopher M.",
    title = "{Dark Atoms: Asymmetry and Direct Detection}",
    eprint = "1105.2073",
    archivePrefix = "arXiv",
    primaryClass = "hep-ph",
    reportNumber = "FERMILAB-PUB-11-207-T",
    doi = "10.1088/1475-7516/2011/10/011",
    journal = "JCAP",
    volume = "10",
    pages = "011",
    year = "2011"
}

@article{Sommerfeld:1931qaf,
    author = "Sommerfeld, A.",
    title = {{{\"U}ber die Beugung und Bremsung der Elektronen}},
    doi = "10.1002/andp.19314030302",
    journal = "Annalen Phys.",
    volume = "403",
    number = "3",
    pages = "257--330",
    year = "1931"
}

@article{Hisano:2003ec,
    author = "Hisano, Junji and Matsumoto, Shigeki and Nojiri, Mihoko M.",
    title = "{Explosive dark matter annihilation}",
    eprint = "hep-ph/0307216",
    archivePrefix = "arXiv",
    reportNumber = "ICRR-REPORT-500-2003-4, YITP-03-42",
    doi = "10.1103/PhysRevLett.92.031303",
    journal = "Phys. Rev. Lett.",
    volume = "92",
    pages = "031303",
    year = "2004"
}

@article{Hisano:2004ds,
    author = "Hisano, Junji and Matsumoto, Shigeki. and Nojiri, Mihoko M. and Saito, Osamu",
    title = "{Non-perturbative effect on dark matter annihilation and gamma ray signature from galactic center}",
    eprint = "hep-ph/0412403",
    archivePrefix = "arXiv",
    reportNumber = "ICRR-REPORT-513-2004-11, YITP-04-73",
    doi = "10.1103/PhysRevD.71.063528",
    journal = "Phys. Rev. D",
    volume = "71",
    pages = "063528",
    year = "2005"
}

@article{Graesser:2011wi,
    author = "Graesser, Michael L. and Shoemaker, Ian M. and Vecchi, Luca",
    title = "{Asymmetric WIMP dark matter}",
    eprint = "1103.2771",
    archivePrefix = "arXiv",
    primaryClass = "hep-ph",
    reportNumber = "LA-UR-11-00565",
    doi = "10.1007/JHEP10(2011)110",
    journal = "JHEP",
    volume = "10",
    pages = "110",
    year = "2011"
}

@article{Planck:2018vyg,
    author = "Aghanim, N. and others",
    collaboration = "Planck",
    title = "{Planck 2018 results. VI. Cosmological parameters}",
    eprint = "1807.06209",
    archivePrefix = "arXiv",
    primaryClass = "astro-ph.CO",
    doi = "10.1051/0004-6361/201833910",
    journal = "Astron. Astrophys.",
    volume = "641",
    pages = "A6",
    year = "2020",
    note = "[Erratum: Astron.Astrophys. 652, C4 (2021)]"
}

@article{Binder:2017rgn,
    author = "Binder, Tobias and Bringmann, Torsten and Gustafsson, Michael and Hryczuk, Andrzej",
    title = "{Early kinetic decoupling of dark matter: when the standard way of calculating the thermal relic density fails}",
    eprint = "1706.07433",
    archivePrefix = "arXiv",
    primaryClass = "astro-ph.CO",
    doi = "10.1103/PhysRevD.96.115010",
    journal = "Phys. Rev. D",
    volume = "96",
    number = "11",
    pages = "115010",
    year = "2017",
    note = "[Erratum: Phys.Rev.D 101, 099901 (2020)]"
}

@article{Fabbrichesi:2020wbt,
    author = "Fabbrichesi, Marco and Gabrielli, Emidio and Lanfranchi, Gaia",
    title = "{The Dark Photon}",
    eprint = "2005.01515",
    archivePrefix = "arXiv",
    primaryClass = "hep-ph",
    doi = "10.1007/978-3-030-62519-1",
    month = "5",
    year = "2020"
}

@article{Goldstein:2026iuu,
    author = "Goldstein, Samuel and Hill, J. Colin",
    title = "{2{\%} determination of Neff from primordial element abundance, cosmic microwave background, and baryon acoustic oscillation measurements}",
    eprint = "2603.13226",
    archivePrefix = "arXiv",
    primaryClass = "astro-ph.CO",
    doi = "10.1103/5vk8-v1rb",
    journal = "Phys. Rev. D",
    volume = "114",
    number = "2",
    pages = "L021305",
    year = "2026"
}

@article{vonHarling:2014kha,
    author = "von Harling, Benedict and Petraki, Kalliopi",
    title = "{Bound-state formation for thermal relic dark matter and unitarity}",
    eprint = "1407.7874",
    archivePrefix = "arXiv",
    primaryClass = "hep-ph",
    reportNumber = "NIKHEF-2014-018",
    doi = "10.1088/1475-7516/2014/12/033",
    journal = "JCAP",
    volume = "12",
    pages = "033",
    year = "2014"
}

@article{Petraki:2015hla,
    author = "Petraki, Kalliopi and Postma, Marieke and Wiechers, Michael",
    title = "{Dark-matter bound states from Feynman diagrams}",
    eprint = "1505.00109",
    archivePrefix = "arXiv",
    primaryClass = "hep-ph",
    reportNumber = "NIKHEF-2015-013",
    doi = "10.1007/JHEP06(2015)128",
    journal = "JHEP",
    volume = "06",
    pages = "128",
    year = "2015"
}

@article{Cyr-Racine:2012tfp,
    author = "Cyr-Racine, Francis-Yan and Sigurdson, Kris",
    title = "{Cosmology of atomic dark matter}",
    eprint = "1209.5752",
    archivePrefix = "arXiv",
    primaryClass = "astro-ph.CO",
    doi = "10.1103/PhysRevD.87.103515",
    journal = "Phys. Rev. D",
    volume = "87",
    number = "10",
    pages = "103515",
    year = "2013"
}

@article{Bansal:2022qbi,
    author = "Bansal, Saurabh and Barron, Jared and Curtin, David and Tsai, Yuhsin",
    title = "{Precision cosmological constraints on atomic dark matter}",
    eprint = "2212.02487",
    archivePrefix = "arXiv",
    primaryClass = "hep-ph",
    doi = "10.1007/JHEP10(2023)095",
    journal = "JHEP",
    volume = "10",
    pages = "095",
    year = "2023"
}

@article{Barron:2026nks,
    author = "Barron, Jared and Essig, Rouven and McDuffie, Megan H. and P{\'e}rez-R{\'\i}os, Jes{\'u}s and Suczewski, Gregory",
    title = "{Pushing the limits of atomic dark matter: First-principles recombination rates and cosmological constraints}",
    eprint = "2602.10197",
    archivePrefix = "arXiv",
    primaryClass = "hep-ph",
    reportNumber = "YITP-SB-2026-03",
    doi = "10.1088/1475-7516/2026/07/090",
    journal = "JCAP",
    volume = "07",
    pages = "090",
    year = "2026"
}

@article{Feng:2009mn,
    author = "Feng, Jonathan L. and Kaplinghat, Manoj and Tu, Huitzu and Yu, Hai-Bo",
    title = "{Hidden Charged Dark Matter}",
    eprint = "0905.3039",
    archivePrefix = "arXiv",
    primaryClass = "hep-ph",
    reportNumber = "UCI-TR-2009-06",
    doi = "10.1088/1475-7516/2009/07/004",
    journal = "JCAP",
    volume = "07",
    pages = "004",
    year = "2009"
}

@article{Agrawal:2016quu,
    author = "Agrawal, Prateek and Cyr-Racine, Francis-Yan and Randall, Lisa and Scholtz, Jakub",
    title = "{Make Dark Matter Charged Again}",
    eprint = "1610.04611",
    archivePrefix = "arXiv",
    primaryClass = "hep-ph",
    doi = "10.1088/1475-7516/2017/05/022",
    journal = "JCAP",
    volume = "05",
    pages = "022",
    year = "2017"
}

@article{Petraki:2014uza,
    author = "Petraki, Kalliopi and Pearce, Lauren and Kusenko, Alexander",
    title = "{Self-interacting asymmetric dark matter coupled to a light massive dark photon}",
    eprint = "1403.1077",
    archivePrefix = "arXiv",
    primaryClass = "hep-ph",
    reportNumber = "NIKHEF-2014-007",
    doi = "10.1088/1475-7516/2014/07/039",
    journal = "JCAP",
    volume = "07",
    pages = "039",
    year = "2014"
}

@article{Fan:2013tia,
    author = "Fan, JiJi and Katz, Andrey and Randall, Lisa and Reece, Matthew",
    title = "{Dark-Disk Universe}",
    eprint = "1303.3271",
    archivePrefix = "arXiv",
    primaryClass = "hep-ph",
    doi = "10.1103/PhysRevLett.110.211302",
    journal = "Phys. Rev. Lett.",
    volume = "110",
    number = "21",
    pages = "211302",
    year = "2013"
}

@article{Rosenberg:2017qia,
    author = "Rosenberg, Eliott and Fan, JiJi",
    title = "{Cooling in a Dissipative Dark Sector}",
    eprint = "1705.10341",
    archivePrefix = "arXiv",
    primaryClass = "astro-ph.GA",
    doi = "10.1103/PhysRevD.96.123001",
    journal = "Phys. Rev. D",
    volume = "96",
    number = "12",
    pages = "123001",
    year = "2017"
}

@article{Ghalsasi:2017jna,
    author = "Ghalsasi, Akshay and McQuinn, Matthew",
    title = "{Exploring the astrophysics of dark atoms}",
    eprint = "1712.04779",
    archivePrefix = "arXiv",
    primaryClass = "astro-ph.GA",
    doi = "10.1103/PhysRevD.97.123018",
    journal = "Phys. Rev. D",
    volume = "97",
    number = "12",
    pages = "123018",
    year = "2018"
}

@article{Roy:2023zar,
    author = "Roy, Sandip and Shen, Xuejian and Lisanti, Mariangela and Curtin, David and Murray, Norman and Hopkins, Philip F.",
    title = "{Simulating Atomic Dark Matter in Milky Way Analogs}",
    eprint = "2304.09878",
    archivePrefix = "arXiv",
    primaryClass = "astro-ph.GA",
    doi = "10.3847/2041-8213/ace2c8",
    journal = "Astrophys. J. Lett.",
    volume = "954",
    number = "2",
    pages = "L40",
    year = "2023"
}

@article{Peebles:1968ja,
    author = "Peebles, P. J. E.",
    title = "{Recombination of the Primeval Plasma}",
    doi = "10.1086/149628",
    journal = "Astrophys. J.",
    volume = "153",
    pages = "1",
    year = "1968"
}

@article{SimonsObservatory:2018koc,
    author = "Ade, Peter and others",
    collaboration = "Simons Observatory",
    title = "{The Simons Observatory: Science goals and forecasts}",
    eprint = "1808.07445",
    archivePrefix = "arXiv",
    primaryClass = "astro-ph.CO",
    doi = "10.1088/1475-7516/2019/02/056",
    journal = "JCAP",
    volume = "02",
    pages = "056",
    year = "2019"
}

@article{Abazajian:2019eic,
    author = "Abazajian, Kevork and others",
    title = "{CMB-S4 Science Case, Reference Design, and Project Plan}",
    eprint = "1907.04473",
    archivePrefix = "arXiv",
    primaryClass = "astro-ph.IM",
    reportNumber = "FERMILAB-PUB-19-431-AE-SCD",
    month = "7",
    year = "2019"
}

@article{NASAPICO:2019thw,
    author = "Hanany, Shaul and others",
    collaboration = "NASA PICO",
    title = "{PICO: Probe of Inflation and Cosmic Origins}",
    eprint = "1902.10541",
    archivePrefix = "arXiv",
    primaryClass = "astro-ph.IM",
    month = "3",
    year = "2019"
}

@article{MacInnis:2023vif,
    author = "MacInnis, Amanda and Sehgal, Neelima and Rothermel, Miriam",
    title = "{Cosmological parameter forecasts for a CMB-HD survey}",
    eprint = "2309.03021",
    archivePrefix = "arXiv",
    primaryClass = "astro-ph.CO",
    doi = "10.1103/PhysRevD.109.063527",
    journal = "Phys. Rev. D",
    volume = "109",
    number = "6",
    pages = "063527",
    year = "2024"
}

@article{Sehgal:2019ewc,
    author = "Sehgal, Neelima and others",
    title = "{CMB-HD: An Ultra-Deep, High-Resolution Millimeter-Wave Survey Over Half the Sky}",
    eprint = "1906.10134",
    archivePrefix = "arXiv",
    primaryClass = "astro-ph.CO",
    journal = "Bull. Am. Astron. Soc.",
    volume = "51",
    number = "7",
    pages = "1--23",
    year = "2019"
}

@article{CMB-HD:2022bsz,
    author = "Aiola, Simone and others",
    collaboration = "CMB-HD",
    title = "{Snowmass2021 CMB-HD White Paper}",
    eprint = "2203.05728",
    archivePrefix = "arXiv",
    primaryClass = "astro-ph.CO",
    reportNumber = "FERMILAB-PUB-22-344-PPD",
    month = "3",
    year = "2022"
}

@article{Sakurai:1960ju,
    author = "Sakurai, J. J.",
    title = "{Theory of strong interactions}",
    doi = "10.1016/0003-4916(60)90126-3",
    journal = "Annals Phys.",
    volume = "11",
    pages = "1--48",
    year = "1960"
}

@article{Kroll:1967it,
    author = "Kroll, N. M. and Lee, T. D. and Zumino, B.",
    editor = "Feinberg, G.",
    title = "{Neutral Vector Mesons and the Hadronic Electromagnetic Current}",
    doi = "10.1103/PhysRev.157.1376",
    journal = "Phys. Rev.",
    volume = "157",
    pages = "1376--1399",
    year = "1967"
}

@article{Fukuda:1974kn,
    author = "Fukuda, Takashi and Kawasaki, Mamoru and Yanagida, Tsutomu and Yonezawa, Minoru",
    title = "{Pi eta Mixing}",
    reportNumber = "RRK 74-4",
    doi = "10.1143/PTP.53.1135",
    journal = "Prog. Theor. Phys.",
    volume = "53",
    pages = "1135",
    year = "1975"
}

@article{Kamada:2019jch,
    author = "Kamada, Ayuki and Yamada, Masaki and Yanagida, Tsutomu T.",
    title = "{Unification for darkly charged dark matter}",
    eprint = "1908.00207",
    archivePrefix = "arXiv",
    primaryClass = "hep-ph",
    doi = "10.1103/PhysRevD.102.015012",
    journal = "Phys. Rev. D",
    volume = "102",
    number = "1",
    pages = "015012",
    year = "2020"
}

@article{Gherghetta:2019coi,
    author = {Gherghetta, Tony and Kersten, J{\"o}rn and Olive, Keith and Pospelov, Maxim},
    title = "{Evaluating the price of tiny kinetic mixing}",
    eprint = "1909.00696",
    archivePrefix = "arXiv",
    primaryClass = "hep-ph",
    reportNumber = "FTPI-MINN-19/23, UMN-TH-3832/19",
    doi = "10.1103/PhysRevD.100.095001",
    journal = "Phys. Rev. D",
    volume = "100",
    number = "9",
    pages = "095001",
    year = "2019"
}

@article{Buckley:2011ye,
    author = "Buckley, Matthew R. and Profumo, Stefano",
    title = "{Regenerating a Symmetry in Asymmetric Dark Matter}",
    eprint = "1109.2164",
    archivePrefix = "arXiv",
    primaryClass = "hep-ph",
    reportNumber = "FERMILAB-PUB-11-437-A",
    doi = "10.1103/PhysRevLett.108.011301",
    journal = "Phys. Rev. Lett.",
    volume = "108",
    pages = "011301",
    year = "2012"
}

@article{Cirelli:2011ac,
    author = "Cirelli, Marco and Panci, Paolo and Servant, Geraldine and Zaharijas, Gabrijela",
    title = "{Consequences of DM/antiDM Oscillations for Asymmetric WIMP Dark Matter}",
    eprint = "1110.3809",
    archivePrefix = "arXiv",
    primaryClass = "hep-ph",
    reportNumber = "CERN-PH-TH-2011-045, SACLAY-T11-022, CP3-ORIGINS-2011-031, DIAS-2011-23",
    doi = "10.1088/1475-7516/2012/03/015",
    journal = "JCAP",
    volume = "03",
    pages = "015",
    year = "2012"
}

@article{Tulin:2012re,
    author = "Tulin, Sean and Yu, Hai-Bo and Zurek, Kathryn M.",
    title = "{Oscillating Asymmetric Dark Matter}",
    eprint = "1202.0283",
    archivePrefix = "arXiv",
    primaryClass = "hep-ph",
    reportNumber = "MCTP-12-03",
    doi = "10.1088/1475-7516/2012/05/013",
    journal = "JCAP",
    volume = "05",
    pages = "013",
    year = "2012"
}

@article{Okada:2012rm,
    author = "Okada, Nobuchika and Seto, Osamu",
    title = "{Originally Asymmetric Dark Matter}",
    eprint = "1205.2844",
    archivePrefix = "arXiv",
    primaryClass = "hep-ph",
    reportNumber = "HGU-CAP-16",
    doi = "10.1103/PhysRevD.86.063525",
    journal = "Phys. Rev. D",
    volume = "86",
    pages = "063525",
    year = "2012"
}

@article{Hut:1979xw,
    author = "Hut, P. and Olive, Keith A.",
    title = "{A COSMOLOGICAL UPPER LIMIT ON THE MASS OF HEAVY NEUTRINOS}",
    doi = "10.1016/0370-2693(79)90039-X",
    journal = "Phys. Lett. B",
    volume = "87",
    pages = "144--146",
    year = "1979"
}

@article{Nussinov:1985xr,
    author = "Nussinov, S.",
    title = "{TECHNOCOSMOLOGY: COULD A TECHNIBARYON EXCESS PROVIDE A 'NATURAL' MISSING MASS CANDIDATE?}",
    reportNumber = "CLNS-85/703",
    doi = "10.1016/0370-2693(85)90689-6",
    journal = "Phys. Lett. B",
    volume = "165",
    pages = "55--58",
    year = "1985"
}

@article{Gelmini:1986zz,
    author = "Gelmini, G. B. and Hall, Lawrence J. and Lin, M. J.",
    title = "{What Is the Cosmion?}",
    reportNumber = "HUTP-86-A042",
    doi = "10.1016/0550-3213(87)90424-X",
    journal = "Nucl. Phys. B",
    volume = "281",
    pages = "726",
    year = "1987"
}

@article{Barr:1990ca,
    author = "Barr, Stephen M. and Chivukula, R. Sekhar and Farhi, Edward",
    title = "{Electroweak Fermion Number Violation and the Production of Stable Particles in the Early Universe}",
    reportNumber = "NSF-ITP-90-27, BA-90-7, BUHEP-90-6, MIT-CTP-1833",
    doi = "10.1016/0370-2693(90)91661-T",
    journal = "Phys. Lett. B",
    volume = "241",
    pages = "387--391",
    year = "1990"
}

@article{Barr:1991qn,
    author = "Barr, Stephen M.",
    title = "{Baryogenesis, sphalerons and the cogeneration of dark matter}",
    reportNumber = "BA-91-12",
    doi = "10.1103/PhysRevD.44.3062",
    journal = "Phys. Rev. D",
    volume = "44",
    pages = "3062--3066",
    year = "1991"
}

@article{Kaplan:1991ah,
    author = "Kaplan, David B.",
    title = "{A Single explanation for both the baryon and dark matter densities}",
    reportNumber = "UCSD-PTH-91-22",
    doi = "10.1103/PhysRevLett.68.741",
    journal = "Phys. Rev. Lett.",
    volume = "68",
    pages = "741--743",
    year = "1992"
}

@article{Dodelson:1991iv,
    author = "Dodelson, Scott and Greene, Brian R. and Widrow, Lawrence M.",
    title = "{Baryogenesis, dark matter and the width of the Z}",
    reportNumber = "HUTP-90-A070, CLNS-90-1032",
    doi = "10.1016/0550-3213(92)90328-9",
    journal = "Nucl. Phys. B",
    volume = "372",
    pages = "467--493",
    year = "1992"
}

@article{Kuzmin:1996he,
    author = "Kuzmin, Vadim A.",
    editor = "Ball, S. J. and Kamyshkov, Y. A.",
    title = "{A Simultaneous solution to baryogenesis and dark matter problems}",
    eprint = "hep-ph/9701269",
    archivePrefix = "arXiv",
    doi = "10.1134/1.953070",
    journal = "Phys. Part. Nucl.",
    volume = "29",
    pages = "257--265",
    year = "1998"
}

@article{Hooper:2004dc,
    author = "Hooper, Dan and March-Russell, John and West, Stephen M.",
    title = "{Asymmetric sneutrino dark matter and the Omega(b) / Omega(DM) puzzle}",
    eprint = "hep-ph/0410114",
    archivePrefix = "arXiv",
    reportNumber = "OUTP-04-17P",
    doi = "10.1016/j.physletb.2004.11.047",
    journal = "Phys. Lett. B",
    volume = "605",
    pages = "228--236",
    year = "2005"
}

@article{Kitano:2004sv,
    author = "Kitano, Ryuichiro and Low, Ian",
    title = "{Dark matter from baryon asymmetry}",
    eprint = "hep-ph/0411133",
    archivePrefix = "arXiv",
    doi = "10.1103/PhysRevD.71.023510",
    journal = "Phys. Rev. D",
    volume = "71",
    pages = "023510",
    year = "2005"
}

@article{Kitano:2005ge,
    author = "Kitano, Ryuichiro and Low, Ian",
    title = "{Grand unification, dark matter, baryon asymmetry, and the small scale structure of the universe}",
    eprint = "hep-ph/0503112",
    archivePrefix = "arXiv",
    month = "3",
    year = "2005"
}

@article{Farrar:2005zd,
    author = "Farrar, Glennys R. and Zaharijas, Gabrijela",
    title = "{Dark matter and the baryon asymmetry}",
    eprint = "hep-ph/0510079",
    archivePrefix = "arXiv",
    doi = "10.1103/PhysRevLett.96.041302",
    journal = "Phys. Rev. Lett.",
    volume = "96",
    pages = "041302",
    year = "2006"
}

@article{Kaplan:2009ag,
    author = "Kaplan, David E. and Luty, Markus A. and Zurek, Kathryn M.",
    title = "{Asymmetric Dark Matter}",
    eprint = "0901.4117",
    archivePrefix = "arXiv",
    primaryClass = "hep-ph",
    reportNumber = "FERMILAB-PUB-09-345-A-T",
    doi = "10.1103/PhysRevD.79.115016",
    journal = "Phys. Rev. D",
    volume = "79",
    pages = "115016",
    year = "2009"
}

@article{Baumann:2017gkg,
    author = "Baumann, Daniel and Green, Daniel and Wallisch, Benjamin",
    title = "{Searching for light relics with large-scale structure}",
    eprint = "1712.08067",
    archivePrefix = "arXiv",
    primaryClass = "astro-ph.CO",
    doi = "10.1088/1475-7516/2018/08/029",
    journal = "JCAP",
    volume = "08",
    pages = "029",
    year = "2018"
}

@article{Qiu:2023bbp,
    author = "Qiu, Yu-Cheng and Sheng, Jie and Tan, Liang and Xing, Chuan-Yang",
    title = "{Ultraheavy atomic dark matter freeze-out through rearrangement}",
    eprint = "2312.13758",
    archivePrefix = "arXiv",
    primaryClass = "hep-ph",
    doi = "10.1103/PhysRevD.109.095007",
    journal = "Phys. Rev. D",
    volume = "109",
    number = "9",
    pages = "095007",
    year = "2024"
}

@article{Ibe:2019yra,
    author = "Ibe, Masahiro and Kobayashi, Shin and Nagai, Ryo and Nakano, Wakutaka",
    title = "{Oscillating Composite Asymmetric Dark Matter}",
    eprint = "1907.11464",
    archivePrefix = "arXiv",
    primaryClass = "hep-ph",
    reportNumber = "IPMU 19-0103",
    doi = "10.1007/JHEP01(2020)027",
    journal = "JHEP",
    volume = "01",
    pages = "027",
    year = "2020"
}

@article{Nagata:2026pbj,
    author = "Nagata, Natsumi and Yanagida, Tsutomu T.",
    title = "{Asymmetric Inelastic Dark Matter and the LUX-ZEPLIN event}",
    eprint = "2609.18564",
    archivePrefix = "arXiv",
    primaryClass = "hep-ph",
    month = "9",
    year = "2026"
}

@article{Pearce:2013ola,
    author = "Pearce, Lauren and Kusenko, Alexander",
    title = "{Indirect Detection of Self-Interacting Asymmetric Dark Matter}",
    eprint = "1303.7294",
    archivePrefix = "arXiv",
    primaryClass = "hep-ph",
    reportNumber = "IPMU13-0137",
    doi = "10.1103/PhysRevD.87.123531",
    journal = "Phys. Rev. D",
    volume = "87",
    pages = "123531",
    year = "2013"
}

@article{Pearce:2015zca,
    author = "Pearce, Lauren and Petraki, Kalliopi and Kusenko, Alexander",
    title = "{Signals from dark atom formation in halos}",
    eprint = "1502.01755",
    archivePrefix = "arXiv",
    primaryClass = "hep-ph",
    reportNumber = "NIKHEF-2015-003, FTPI-MINN-15-05",
    doi = "10.1103/PhysRevD.91.083532",
    journal = "Phys. Rev. D",
    volume = "91",
    pages = "083532",
    year = "2015"
}

@article{Baldes:2017gzw,
    author = "Baldes, Iason and Petraki, Kalliopi",
    title = "{Asymmetric thermal-relic dark matter: Sommerfeld-enhanced freeze-out, annihilation signals and unitarity bounds}",
    eprint = "1703.00478",
    archivePrefix = "arXiv",
    primaryClass = "hep-ph",
    reportNumber = "DESY-17-034, NIKHEF-2017-009",
    doi = "10.1088/1475-7516/2017/09/028",
    journal = "JCAP",
    volume = "09",
    pages = "028",
    year = "2017"
}

@article{Khlopov:2010ik,
    author = "Khlopov, Maxim Yu. and Mayorov, Andrey G. and Soldatov, Evgeny Yu.",
    title = "{The dark atoms of dark matter}",
    eprint = "1012.0934",
    archivePrefix = "arXiv",
    primaryClass = "astro-ph.CO",
    journal = "Prespace. J.",
    volume = "1",
    pages = "1403--1417",
    year = "2010"
}

@article{Ibe:2011hq,
    author = "Ibe, Masahiro and Matsumoto, Shigeki and Yanagida, Tsutomu T.",
    title = "{The GeV-scale dark matter with B{\textendash}L asymmetry}",
    eprint = "1110.5452",
    archivePrefix = "arXiv",
    primaryClass = "hep-ph",
    doi = "10.1016/j.physletb.2012.01.032",
    journal = "Phys. Lett. B",
    volume = "708",
    pages = "112--118",
    year = "2012"
}

@article{Nagata:2016knk,
    author = "Nagata, Natsumi and Olive, Keith A. and Zheng, Jiaming",
    title = "{Asymmetric Dark Matter Models in SO(10)}",
    eprint = "1611.04693",
    archivePrefix = "arXiv",
    primaryClass = "hep-ph",
    reportNumber = "UT--16--31, UMN--TH--3612-16, FTPI--MINN--16-32",
    doi = "10.1088/1475-7516/2017/02/016",
    journal = "JCAP",
    volume = "02",
    pages = "016",
    year = "2017"
}

@article{An:2009vq,
    author = "An, Haipeng and Chen, Shao-Long and Mohapatra, Rabindra N. and Zhang, Yue",
    title = "{Leptogenesis as a Common Origin for Matter and Dark Matter}",
    eprint = "0911.4463",
    archivePrefix = "arXiv",
    primaryClass = "hep-ph",
    reportNumber = "UMD-40762-471, UMD-PP-09-062, IC-2009-090",
    doi = "10.1007/JHEP03(2010)124",
    journal = "JHEP",
    volume = "03",
    pages = "124",
    year = "2010"
}

\end{document}